\documentclass[11pt,a4paper]{article}
\pdfoutput=1
\usepackage{jheppub}

\usepackage{multirow}
\usepackage{amsmath,amssymb,amsfonts}
\usepackage{graphicx}
\usepackage{color}
\usepackage{bbm}
\usepackage{url}
\usepackage{slashed}
\usepackage{subfigure}
\usepackage[usenames,dvipsnames]{xcolor}
\usepackage{float} 
\usepackage{epsfig}
\usepackage{graphicx}
\usepackage{euscript}
\usepackage{color}
\usepackage{slashed}
\usepackage{epstopdf}
\usepackage{mathbbol}
\usepackage{mathrsfs}
\usepackage[normalem]{ulem}
\usepackage{pifont}
\usepackage{bbold}
\usepackage{float}
\allowdisplaybreaks
\usepackage{bbding}

\usepackage{verbatim}
\usepackage{graphicx}
\usepackage{latexsym}

\newcommand{\cR}{{\mathcal R}}

\newcommand{\be}{\begin{equation}}
\newcommand{\ee}{\end{equation}}
\newcommand{\bea}{\begin{align}}
\newcommand{\eea}{\end{align}}

\newcommand{\gb}{\bar{g}} 
\newcommand{\Db}{\bar{D}} 
\newcommand{\Rb}{\bar{R}} 

\newcommand{\p}{\partial}

\newcommand{\cO}{\mathcal{O}}

\newcommand{\ver}{\mathbb{V}}
\newcommand{\prop}{\mathbb{P}}
\newcommand{\gprop}{\mathbb{G}}

\newcommand{\ct}{\tilde{c}}

\newcommand{\unit}{\mathbbm 1}

\definecolor{darkgreen}{rgb}{0.2,0.6,0}
\definecolor{lightblue}{rgb}{0,0.5,0.8}
\definecolor{lightred}{rgb}{0.8,0.2,0.2}
\definecolor{darkorange}{rgb}{1,0.549,0}
\definecolor{brown}{rgb}{0.609, 0.164, 0.164}

\allowdisplaybreaks

\title{
Scalar-Tensor theories within Asymptotic Safety}
\author[a]{Cristobal Laporte,} \emailAdd{cristobal.laportemunoz@ru.nl}
\affiliation[a]{Institute for Mathematics, Astrophysics and Particle Physics (IMAPP),
Radboud University, Heyendaalseweg 135, 6525 AJ Nijmegen, The Netherlands}

\author[a,b]{Antonio D. Pereira,} \emailAdd{a.duarte@science.ru.nl}
\affiliation[b]{Instituto de F\'isica, Universidade Federal Fluminense, Campus da Praia Vermelha, Av. Litor\^anea s/n, 24210-346, Niter\'oi, RJ, Brazil}

\author[a]{Frank Saueressig,} \emailAdd{f.saueressig@science.ru.nl}

\author[a]{Jian Wang} \emailAdd{jian.wang@science.ru.nl}

\abstract{
Asymptotic Safety provides an elegant mechanism for obtaining a consistent high-energy completion of gravity and gravity-matter systems. Following the initial idea by Steven Weinberg, the construction builds on an interacting fixed point of the theories renormalization group (RG) flow. In this work we use the Wetterich equation for the effective average action to investigate the RG flow of gravity supplemented by a real scalar field. We give a non-perturbative proof that the subspace of interactions respecting the global shift-symmetry of the scalar kinetic term is closed under RG transformations. Subsequently, we compute the beta functions in an approximation comprising the Einstein-Hilbert action supplemented by the shift-symmetric quartic scalar self-interaction and the two lowest order shift-symmetric interactions coupling scalar-bilinears to the spacetime curvature. The computation utilizes the background field method with an arbitrary background, demonstrating that the results are manifestly background independent. Our beta functions exhibit an interacting fixed point suitable for Asymptotic Safety, where all matter interactions are non-vanishing. The presence of this fixed point is rooted in the interplay of the matter couplings which our work tracks for the first time. The relation of our findings with previous results in the literature is discussed in detail and we conclude with a brief outlook on potential phenomenological applications.
}

\begin{document}
\maketitle
\section{Introduction}
\label{sect.Intro}

The structure of spacetime at very short length scales remains unknown. It is expected that a description of this regime requires a theory of quantum gravity since quantum fluctuations of spacetime itself become relevant as one zooms in on its microstructure and substantial deviations from a smooth manifold structure can take place. Different approaches to formulate a theory of quantum gravity are based on very different theoretical assumptions and can lead to distinct physical pictures of quantum spacetime.  Irrespective of its details, the microscopic picture has to give rise to a description of spacetime in terms of metric degrees of freedom as one flows to macroscopic scales. Thus, it is suggestive to construct a quantum theory of gravity in terms of metric variables directly. Starting out from the Einstein-Hilbert action, this leads to a perturbatively non-renormalizable quantum field theory \cite{tHooft:1974toh,Christensen:1979iy,Goroff:1985th}: absorbing the infinities arising in the quantization procedure requires infinitely many counterterms, each one introducing a new free parameter into the theory. Since each free parameter has to be fixed by experimental input, this leads to a breakdown of predictivity of the underlying quantum field theory. This does not reflect an incompatibility between the standard framework of continuum quantum field theory and gravity, however. In practice, quantum-gravity corrections can be computed explicitly in an effective field theory framework which is reliable for energy scales way below the Planck mass \cite{Donoghue:1993eb,Donoghue:1994dn,Burgess:2003jk}. Insisting on a perturbative treatment, one can improve the renormalizability properties by adding higher-derivative terms \cite{Stelle:1976gc}. This results in a perturbatively renormalizable theory, but challenges perturbative unitarity or microcausality, see, e.g., \cite{Modesto:2015ozb,Anselmi:2017ygm,Anselmi:2018ibi,Anselmi:2018tmf,Donoghue:2019fcb,Donoghue:2019ecz} for recent discussions.

Yet, a quantum field theory of gravity based on metric variables may exist as a non-perturbative theory. If the couplings of the full effective action reach an ultraviolet (UV) fixed point, then physical quantities constructed out of those couplings will be well-behaved in the deep ultraviolet due to quantum scale invariance \cite{Wetterich:2019qzx}. The enhanced symmetry is provided by an interacting renormalization group (RG) fixed point, a so-called non-Gaussian fixed point (NGFP). The fixed point ensures that the dimensionless versions of the couplings retain finite values up to arbitrarily high energy scales, which in turn ensures that dimensionless observables like scattering amplitudes remain finite. Such a situation is well-known in Yang-Mills theory where such a fixed point sits at vanishing values of the gauge coupling leading to asymptotic freedom. In this case, perturbation theory is applicable in the vicinity of the fixed point. Conversely, in quantum gravity, following the idea put forward by Weinberg in \cite{Hawking:1979ig}, the underlying quantum field theory might be asymptotically safe, i.e., there is a fixed point where the couplings (or, at least, a subset of couplings) do not vanish. Substantial progress in testing Weinberg's proposal in an Euclidean setting was achieved after the pioneering work by Reuter \cite{Reuter:1996cp}, where the functional renormalization group (FRG) for the effective average action $\Gamma_k$ \cite{Wetterich:1992yh,Morris:1993qb,Reuter:1993kw} was adapted to quantum gravity. For reviews on the FRG, we refer to \cite{Berges:2000ew,Pawlowski:2005xe,Gies:2006wv,Dupuis:2020fhh}. Being non-perturbative, the FRG allows for computations beyond  expansions in a small parameter, although each concrete, non-perturbative computation requires a proper truncation of  $\Gamma_k$. Within truncations, compelling evidence for the existence of a non-trivial fixed point in pure (Euclidean) gravity was obtained over the past two decades \cite{Souma:1999at,Lauscher:2001ya,Reuter:2001ag,Lauscher:2001rz,Lauscher:2002sq,Reuter:2002kd,Niedermaier:2002eq,Litim:2003vp,Codello:2006in,Machado:2007ea,Codello:2007bd,Codello:2008vh,Benedetti:2009rx,Machado:2009ph,Benedetti:2009gn,Niedermaier:2009zz,Manrique:2009uh,Manrique:2010am,Groh:2010ta,Eichhorn:2010tb,Benedetti:2010nr,Manrique:2011jc,Benedetti:2012dx,Christiansen:2012rx,Dietz:2012ic,Becker:2012js,Codello:2013fpa,Ohta:2013uca,Falls:2013bv,Becker:2014pea,Falls:2014tra,Christiansen:2014raa,Becker:2014qya,Demmel:2014hla,Christiansen:2015rva,Morris:2015oca,Ohta:2015efa,Ohta:2015fcu,Gies:2015tca,Biemans:2016rvp,Gies:2016con,Denz:2016qks,Platania:2017djo,Houthoff:2017oam,Falls:2017lst,Knorr:2017fus,Christiansen:2017bsy,Becker:2018quq,DeBrito:2018hur,Falls:2018ylp,Knorr:2018kog,Becker:2019fhi,Becker:2019tlf,Becker:2020mjl,Kluth:2020bdv,Falls:2020qhj,Knorr:2020ckv,Bonanno:2021squ,Martini:2021slj,Knorr:2021niv,Martini:2021lcx,Becker:2021pwo,Baldazzi:2021orb}.\footnote{For applications of the FRG \cite{Wetterich:1992yh,Morris:1993qb,Reuter:1993kw} in the context of theories encoding the metric degrees of freedom in other variables (vierbeins, connections) or implementing different symmetries (Weyl gravity, unimodular gravity, Einstein-Cartan gravity) the reader may consult \cite{Daum:2010qt,Codello:2012sn,Harst:2012ni,Daum:2013fu,Harst:2014vca,Harst:2015eha,DeBrito:2019gdd,deBrito:2020rwu,deBrito:2020xhy,deBrito:2021pmw,Ferrero:2021xqg}. Furthermore, many conceptual questions related to Asymptotic Safety have been addressed in \cite{Lauscher:2005qz,Reuter:2006zq,Manrique:2008zw,Reuter:2008qx,Reuter:2008wj,Reuter:2011ah,Nink:2012vd,Nink:2014yya,Reuter:2015rta,Nink:2015lmq,Ohta:2015zwa,Ohta:2016npm,Pagani:2013fca,Pagani:2015ema,Pagani:2016dof,Pagani:2019vfm,Pagani:2020say}.} Whether such a fixed point exists beyond finite-dimensional truncations and persists in a Lorentzian setting remains an open challenging problem \cite{Donoghue:2019clr,Bonanno:2020bil}, even though the first steps taken in \cite{Bridle:2013sra,Demmel:2014sga,Demmel:2015oqa,Gonzalez-Martin:2017gza} and \cite{Manrique:2011jc,Biemans:2016rvp,Biemans:2017zca,Houthoff:2017oam,Knorr:2018fdu} have yielded encouraging results in these directions already. In this work, we will refer to the quantum theory of the gravitational interactions emerging from this construction as asymptotically safe quantum gravity (ASQG). For reviews on the topic, we refer to \cite{Niedermaier:2006wt,Reuter:2012id,Eichhorn:2017egq,Percacci:2017fkn,Eichhorn:2018yfc,Reuter:2019byg,Pereira:2019dbn,Reichert:2020mja,Pawlowski:2020qer}.

In complete analogy to the fate of asymptotic freedom in gauge theories coupled to matter, a natural concern in this approach is whether the pure-gravity fixed point (also called the Reuter fixed point) has analogues when matter degrees of freedom are included: any realistic description of our world should accommodate, at least, the matter degrees of freedom of the Standard Model of particle physics with possible minimal extensions such as right-handed neutrinos or matter degrees of freedom linked to dark matter and dark energy. Thus, realistic models should include suitable matter degrees of freedom, thereby falling into the class of asymptotically safe quantum gravity-matter (ASQGM) theories. This perspective has been investigated within Asymptotic Safety for a long time \cite{Griguolo:1995db,Dou:1997fg,Percacci:2002ie,Percacci:2003jz,Zanusso:2009bs,Narain:2009fy,Narain:2009gb,Daum:2009dn,Shaposhnikov:2009pv,Daum:2010bc,Harst:2011zx,Folkerts:2011jz,Eichhorn:2011pc,Eichhorn:2012va,Dona:2012am,Dona:2013qba,Dona:2014pla,Labus:2015ska,Meibohm:2015twa,Oda:2015sma,Dona:2015tnf,Percacci:2015wwa,Eichhorn:2016esv,Eichhorn:2016vvy,Wetterich:2016uxm,Christiansen:2017gtg,Biemans:2017zca,Hamada:2017rvn,Christiansen:2017qca,Eichhorn:2017eht,Eichhorn:2017ylw,Eichhorn:2017lry,Becker:2017tcx,Eichhorn:2017sok,Christiansen:2017cxa,Eichhorn:2017als,Alkofer:2018fxj,Gies:2018jnv,Eichhorn:2018whv,Eichhorn:2018akn,Alkofer:2018baq,Eichhorn:2018ydy,Pawlowski:2018ixd,Eichhorn:2018nda,Eichhorn:2019yzm,DeBrito:2019rrh,Knorr:2019atm,Wetterich:2019rsn,Reichert:2019car,Burger:2019upn,Platania:2020knd,Daas:2020dyo,Eichhorn:2020kca,Eichhorn:2020sbo,deBrito:2020dta,Ali:2020znq,Gies:2021upb,deBrito:2021pyi,Eichhorn:2021tsx,Daas:2021abx}. A particular feature of gravity-matter systems is that already the kinetic terms of the matter fields give rise to infinitely many vertices coupling matter to an arbitrary number of gravitons. Hence, even if one starts with minimally coupled matter these vertices generate momentum-dependent self-interactions of matter fields due to quantum corrections containing internal graviton lines \cite{Eichhorn:2012va,Eichhorn:2017eht}. Moreover, non-minimal couplings between curvature structures and (derivatives of) matter fields are not forbidden by the underlying symmetries of the gravity-matter system and, therefore, are expected to appear as well. This leads to a proliferation of matter couplings which will be quantum mechanically generated even if not included in the original bare action of the theory. By now, many results obtained within truncations in the FRG framework point towards the existence of such a fixed point. The challenge for ASQGM-models is to establish that the fixed points seen at minimal coupling survive once the web of non-vanishing gravity-matter interactions is considered. 

Since ASQGM-models build on interacting RG fixed points, predictivity must be established based on the properties of the fixed point. This is at variance with UV completions provided by free fixed points, where the number of relevant directions is equivalent to the number of couplings with a positive or marginally relevant mass dimension.  Generically, Asymptotic Safety will be predictive if the number of directions along which an RG flow can emanate from the fixed point (the number of UV-relevant directions) is finite.
Based on the explicit calculations performed so far, the number of relevant directions seems to saturate at a small value for large truncations \cite{Alkofer:2018fxj}.\footnote{Investigating large-scale truncations within the composite operator formalism suggests that there can also gravity-matter fixed points beyond near-canonical scaling \cite{Houthoff:2020zqy,Kurov:2020csd}. These are difficult to track based on FRG computations though.} This provided a first non-trivial check of the desired properties of the fixed point. Moreover, different results  suggest that the fixed points which currently feature in ASQGM-models exhibit a near-canonical scaling \cite{Falls:2017lst,Falls:2018ylp,Kluth:2020bdv,Eichhorn:2018akn,Eichhorn:2018ydy}, in the sense that the relevance of an operator is still set by its canonical mass dimension. This suggests that canonical power-counting is a reasonable guiding principle for the construction of truncations exploring the predictive power of these models.   

In this work, we perform the currently most advanced investigation of ASQGM systems in the framework of gravity coupled to a single scalar field (scalar-tensor theories). Besides the Einstein-Hilbert action, we include the scalar anomalous dimensions, the two dimension-six operators coupling the spacetime curvature to scalar bilinears, as well as the dimension-eight  momentum-dependent  scalar self-interaction. This selection is guided by a symmetry principle: all operators associated with the gravity-matter interactions are compatible with the $\mathbb{Z}_2$- and global shift symmetry exhibited by the scalar kinetic term. As pointed out in \cite{Eichhorn:2012va,Eichhorn:2017eht}, the corresponding couplings take non-zero values at the fixed point. Testing whether the fixed point found at minimal coupling \cite{Dona:2013qba} survives once the non-minimal terms are included provides an acid-test for the consistency of the ASQGM-model. 
	
Notably, our work includes the recent investigations \cite{Eichhorn:2012va,Eichhorn:2017sok,deBrito:2021pyi} as subcases. Our beta functions provide an overarching umbrella on these works, thereby retaining \emph{all backreactions} of the non-minimal gravity-matter couplings on the corresponding beta functions. The latter are identified to be of essential importance in order for the full system exhibiting an interacting fixed point suitable for Asymptotic Safety. The results are obtained within a background computation. The novel technical feature is that the background used in the computation is left unspecified, i.e., our results are shown to be manifestly \emph{background-independent}. Instead we resort to projection rules which allow to systematically trace and eliminate contributions which do not contribute to the present setting.

The paper is organized as follows. In Sect.~\ref{Sect:OverviewFRG} we provide a brief introduction to the FRG underlying this work and review some arguments behind the non-Gaussian nature of the matter interactions compatible with the global symmetries of the matter kinetic term. In Sect.\ \ref{subsect.3.2},  we show that the global symmetry of the scalar kinetic term implies an exact functional identity that must be respected by the effective average action. Then, in Sect.~\ref{sect.4setup} we specify the details of the truncation employed for $\Gamma_k$, describe the computation of the corresponding beta functions, and give the final result in terms of threshold integrals. Sect.~\ref{Sect.fps} contains our results on the existence and stability properties of fixed points as well as a structural view of the underlying beta functions. We end with our conclusions and an outlook on how our findings can be related to quantum-gravity phenomenology in Sect.\ \ref{Sect.6}. Technical details of our computation as well as lengthy expressions are collected in the appendices.

\section{Asymptotic Safety and the functional renormalization group}
\label{Sect:OverviewFRG}
This section reviews the basics of the functional renormalization group, currently constituting the key technical tool for investigating Asymptotic Safety (Sect.\ \ref{Sect.2.1}) before summarizing the structural aspects of interacting gravity matter fixed points discussed in the literature \cite{Narain:2009fy,Narain:2009gb,Eichhorn:2012va,Eichhorn:2016esv,Eichhorn:2017eht,deBrito:2021pyi} in Sect.\ \ref{Sect.2.2}.
\subsection{The functional renormalization group - a brief introduction}
\label{Sect.2.1}
The FRG provides a smooth implementation of a Wilsonian RG flow, realizing a shell-wise integration of field modes in the Euclidean path integral. This is achieved by the introduction of a regulator function $\Delta S_k$ in the Boltzmann weight of the path integral such that all field modes with momenta smaller than $k$ are suppressed. Lowering $k$ unsuppresses new field modes until all modes have been integrated out at $k=0$. At this point the entire path integral has been performed. As a technical requirement, the regulator term is taken to be of the form
\begin{equation}
\Delta S_k[\varphi] = \frac{1}{2}\int\mathrm{d}^dx\,\sqrt{\gb} \, \varphi(x)\mathcal{R}_k (-\Db^2) \varphi (x)\,,
\label{OFRG1}
\end{equation}
with $\varphi(x)$ representing the collection of fluctuation fields being integrated in the path integral, $d$ is the spacetime dimension, and $-\Db^2$ being the Laplacian constructed from the background spacetime metric $\gb_{\mu\nu}$. The regulator function $\mathcal{R}_k$ is an arbitrary function of the eigenvalues of the Laplacian that must satisfy consistency conditions to be considered a ``proper" regulator though. For eigenvalues $p^2>k^2$, the function $\mathcal{R}_k$ must fall off to zero sufficiently fast, so that modes labelled by such eigenvalues are integrated out without suppression factor. For $p^2<k^2$, the regulator function acts as a mass term, implementing a suppression of these modes. Moreover, $\lim_{k \rightarrow 0} R_k(p^2) =0$, so that the regulator vanishes for $k=0$ and all modes are integrated out. Finally, the regulator should be quadratic in the fluctuation fields in order to act as a mass-type regulator. The last requirement is crucial for defining the effective average action $\Gamma_k$ as a modified Legendre transform of the scale-dependent generating functional of connected correlation functions. The effective average action satisfies a formally exact FRG equation, the Wetterich equation \cite{Wetterich:1992yh,Morris:1993qb,Reuter:1993kw}
\begin{equation}
\partial_t \Gamma_k = \frac{1}{2}\mathrm{STr}\left[\left(\Gamma^{(2)}_k + \mathcal{R}_k\right)^{-1}\partial_t\mathcal{R}_k\right]\,,
\label{OFRG2}
\end{equation}
where $t \equiv \ln(k/k_0)$ denotes the RG time, and $k_0$ is an arbitrary reference momentum scale. 
The supertrace STr represents a sum over all internal indices as well as an integration over spacetime. Moreover, depending on the nature of the field, it introduces a multiplicative factor, i.e., for commuting real fields, this constant is 1, while it is 2 for complex fields. For anti-commuting fields, the constant is -2. Finally, $\Gamma^{(2)}_k$ is the second functional derivative of $\Gamma_k$ with respect to the fluctuation fields which typically is matrix valued in field space. By construction $\Gamma_k$  coincides with  the effective action (one-particle irreducible generating functional) $\Gamma$ as $k=0$. At the UV-cutoff scale $\Lambda$ (which may or may not be taken to infinity) $\Gamma_{\Lambda}$ is equivalent to the bare/microscopic action of the theory $S$. Solving \eqref{OFRG2} and taking the $k\to 0$ is thus equivalent to carrying out the path integral. The simple one-loop structure of \eqref{OFRG2} follows from the quadratic nature of the regulator \eqref{OFRG1}. 

The FRG \eqref{OFRG2} ``lives'' on theory space, built from all action functionals which can be constructed from the field content of the theory and compatible with its symmetries. Formally, $\Gamma_k$ can then be expressed as
\begin{equation}
	\Gamma_k [\phi] = \sum_{i=1}^{\infty}\,\bar{g}_i \,\EuScript{O}_i (\phi)\,,
	\label{OFRG3}
\end{equation}
with $\phi \equiv \langle \varphi\rangle_J$ being the expectation value of the fluctuation field in the presence of external sources $J$. The operators $\EuScript{O}_i (\phi)$ are (local) operators of the fields $\phi(x)$ and $\bar{g}_i$ are the corresponding dimensionful couplings. The $\bar{g}_i$ serve as ``coordinates'' on the theory space and can be traded for their dimensionless counterparts $g_i = \bar{g}_i k^{-d_{g_i}}$ with $d_{g_i}$ being the canonical mass dimension of the coupling $\bar{g}_i$.  From \eqref{OFRG3}, one can write,
\begin{equation}
	\partial_t\Gamma_k [\phi] = \sum_{i=1}^{\infty}\,(\partial_t \bar{g}_i )\,\EuScript{O}_i (\phi) = \sum_{i=1}^{\infty}\,k^{d_{g_i}}(d_{g_i} g_i+\partial_t g_i )\,\EuScript{O}_i (\phi) \equiv \sum_{i=1}^{\infty}\,k^{d_{g_i}}(d_{g_i} g_i+\beta_i )\,\EuScript{O}_i (\phi)\,,
	\label{OFRG4}
\end{equation}
where we have introduced the beta function of the coupling $g_i$
\be\label{eq:betadef}
  \partial_t g_i \equiv \beta_i \, .
\ee
The general structure of the beta functions is
\begin{equation}
	\beta_i = -d_{g_i} \, g_i+ k^{-d_{g_i}} \, \partial_t \bar{g}_i\,,
	\label{OFRG5}
\end{equation}
where the first term is due to the canonical mass dimension of $\bar{g}_i$ and $ k^{-d_{g_i}}\partial_t \bar{g}_i$ is generated by quantum fluctuations. 

At a fixed point $g^{*} = (g^*_1,g^{*}_2,\ldots )$ one has $\beta_i = 0$, $\forall i$. Given a fixed point one can study the linearized flow in its vicinity, encoded in the linearized beta functions
\begin{equation}
	\beta_i (g) = \sum_{j=1}^{\infty} \, B_i{}^j \, (g_j - g^{*}_j) \,.
	\label{OFRG6}
\end{equation}
Here the stability matrix is defined as 
\be\label{def:stabmat}
B_i{}^j \equiv \left. \frac{\partial \beta_i}{\partial g_j}\right|_{g=g^*} \, . 
\ee
Assuming that these objects exist, we denote the right-eigenvectors of $B_i{}^j$ by $V_J$ and the corresponding (potentially complex) eigenvalues by $(-\theta_J)$. The $\theta_J$ are the critical exponents associated with the fixed point. The solution of \eqref{def:stabmat} is then given by
\be\label{linsol}
g_i(t) = g_i^* + \sum_J \, C_J \, V_J^i \, e^{-\theta_J t} \, , 
\ee
where $C_J$ are integration constants. From \eqref{linsol} one readily concludes that the flow along an eigendirection where Re($\theta_J) > 0$ is dragged into the fixed point as $t \rightarrow \infty$. Conversely, eigendirections where Re($\theta_J) < 0$ repell the flow when going to the UV. Thus, in order to approach the fixed point as $t \rightarrow \infty$, the $C_J$ associated with the latter directions must be set to zero. Conversely, the coefficients associated with UV-attractive directions where Re($\theta_J) > 0$ are not fixed by Asymptotic Safety. They constitute free parameters which need to be fixed based on experimental input. This discussion implies that fixed points coming with a lower number of  UV-attractive directions have more predictive power. A consistent proposal for an ASQGM system requires a fixed point which provides the high-energy completion of the quantum theory with (preferably) finitely many UV-attractive directions.

\subsection{Properties of gravity-matter fixed points}
\label{Sect.2.2}
The functional renormalization group methods surveyed in the previous subsection have provided substantial evidence that gravity in four spacetime-dimensions supports a NGFP suitable for Asymptotic Safety. A natural step towards a more realistic theory of quantum gravity related to our world is the inclusion of matter degrees of freedom. At first sight, it is then suggestive that the gravity-matter system should possess an NGFP of the form
\be\label{eq:matterfixedpoints}
\text{NGFP}^{\rm ASQGM} = \text{NGFP}^{\rm ASQG} \otimes \text{GFP}^{\rm matter} \, ,
\ee
i.e., the UV-completion retains interactions in the gravitational sector while the matter sector is non-interacting and minimally coupled. While this structure appears in certain classes of approximations \cite{Narain:2009fy,Narain:2009gb}, e.g., when investigating the existence of gravity-matter fixed points at minimal couplings \cite{Dona:2013qba}, this is an artifact of the approximation: as suggested in \cite{Eichhorn:2012va} and later substantiated in \cite{Eichhorn:2016esv,Eichhorn:2017eht,deBrito:2021pyi}, the interactions in the gravitational sector induce non-minimal interactions as well as matter self-interactions at the $\text{NGFP}^{\rm ASQGM}$. Here we review the underlying argument based on a qualitative analysis.

Our work focuses on gravity, encoded in the spacetime metric $g_{\mu\nu}$ coupled to a real scalar field $\phi$. Writing down the scalar kinetic term,
\be\label{eq:phikin}
\EuScript{I}_{\rm kin} = \frac{Z_k}{2} \int {\rm d}^4x \sqrt{g} \, g^{\mu\nu} (\p_\mu \phi) (\p_\nu \phi) \, , 
\ee
with $Z_k$ being the wave-function renormalization of $\phi$, one easily verifies that $\EuScript{I}_{\rm kin}$ is invariant under the discrete $\mathbb{Z}_2$-symmetry $\phi \mapsto -\phi$ as well as continuous shifts
\be\label{eq:shift}
\phi \to \phi + c \, , \qquad c \in \mathbb{R} \, . 
\ee

\begin{figure}[t]
	\includegraphics[width=\linewidth]{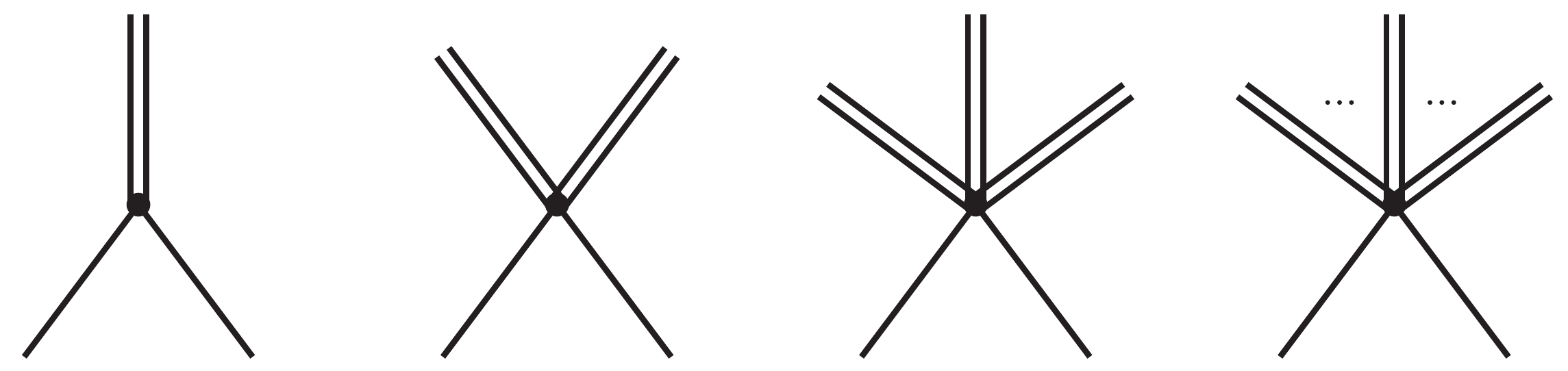}
	\caption{Scalar-graviton vertices generated in the expansion \eqref{GlobalSym1}. Double-lines represent graviton insertions and simple lines denote insertions of the scalar field. Due to the global shift symmetry exhibited by $\EuScript{I}_{\rm kin}$, the above vertices always carry a derivative for each scalar line.}
	\label{sgv1}
\end{figure}
\begin{figure}[t]
\begin{center}
  \includegraphics[scale=0.6]{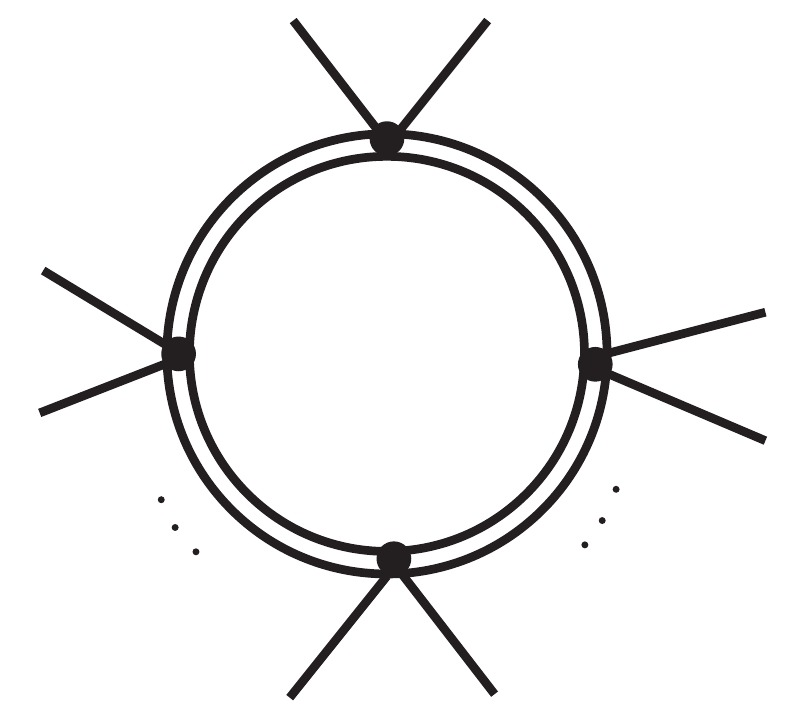}
\end{center}
  \caption{Effective shift-symmetric scalar self-interaction generated by the shift-symmetric vertices depicted in Fig.~\ref{sgv1}.}
  \label{sgv2}
\end{figure}

The kinetic term \eqref{eq:phikin} contains the determinant and the inverse of the metric which generate the infinite tower of gravity-matter vertices illustrated in Fig.\ \ref{sgv1}. Decomposing $g_{\mu\nu}$ in a background metric $\gb_{\mu\nu}$ and fluctuations $h_{\mu\nu}$ using the linear split $g_{\mu\nu} = \gb_{\mu\nu} + h_{\mu\nu}$, this expansion takes the schematic form
\begin{equation}
	 \EuScript{I}_{\rm kin} \sim \frac{Z_k}{2} \int\mathrm{d}^4x\sqrt{\bar{g}}\,\Big(\bar{g}^{\mu\nu}+h^{(1)\mu\nu}+h^{(2)\mu\nu}+\EuScript{O}(h^3)^{\mu\nu}\Big)\partial_\mu \phi\, \partial_\nu \phi\,,
	\label{GlobalSym1}
\end{equation}
where $h^{(n)\mu\nu}$ contains $n$ powers of the fluctuation field $h_{\mu\nu}$. These vertices can be connected by internal graviton lines leading to matter self-interactions, see Fig.~\ref{sgv2}, and non-minimal gravity-matter interactions. All vertices carry the global symmetries of $\EuScript{I}_{\rm kin}$, though. As a consequence, the induced interaction momomials also respect these symmetries \cite{Eichhorn:2017eht}. 
A consequence of this analysis is that the mechanism outlined above will not generate a scalar potential $V(\phi)$ since this is not compatible with the shift-symmetry \eqref{eq:shift}.

\begin{figure}[t]
	\begin{center}
		\includegraphics[width=\linewidth]{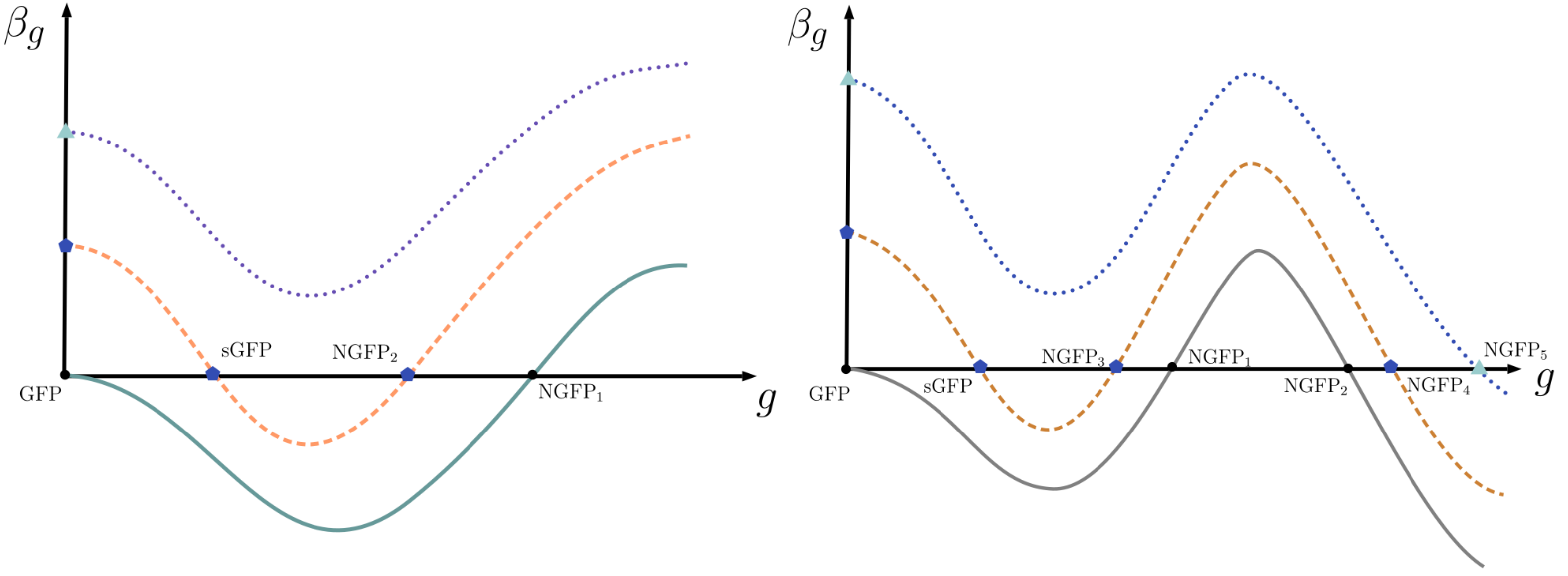}
	\end{center}
	\caption{Illustration of the qualitative structure encountered in the beta functions linked to interactions sharing the same global symmetries as the matter kinetic term. We distinguish two situations: on the left, the beta function is given by a polynomial of degree two in the matter coupling $g$ while, on the right, it has the structure of a cubic. In the absence of gravity (solid gray line), the matter theory has a GFP and, eventually, additional NGFPs, represented by $\mathrm{NGFP}_1$ in the left diagram and $\mathrm{NGFP}_1$ and $\mathrm{NGFP}_2$ in the right diagram. When the matter theory is coupled to gravity, $\beta_g$ receives contributions from the vertices generated by the kinetic term which are independent of the coupling $g$. These contributions eliminate the GFP and shift its position to a finite value of $g$ which balances the $g$-dependent and independent contributions. This generates the shifted Gaussian fixed point (sGFP). In addition, the system can again feature NGFPs represented by $\mathrm{NGFP}_2$ in the left panel, and $\mathrm{NGFP}_{3,4}$ in the right pannel. If quantum-gravity fluctuations are sufficiently strong, the shift in $\beta_g$ represented in the plot on the left may lead to a situation where the fixed points become complex and the real fixed points cease to exist. In this case an asymptotically safe theory of gravity and matter can no longer be realized. This mechanism is operative for beta functions of even order in the coupling only: the cubic beta function depicted on the right pannel, always has a real fixed point $\mathrm{NGFP}_5$ even if the other two fixed points forming the triplet (sGFP, NGFP$_3$, NGFP$_4$) turn into complex roots.}
	\label{sgv3}
\end{figure}

As pointed out in \cite{Eichhorn:2012va,Eichhorn:2017eht}, the couplings associated with interactions preserving the shift symmetry must generically take a non-zero fixed point value. Hence these couplings are not compatible with a Gaussian fixed point in the matter sector and thus do not follow the structure suggested in \eqref{eq:matterfixedpoints}. Let us illustrate this feature based on the momentum-dependent quartic self-interaction 
\begin{equation}
	\EuScript{I}_5 = Z_k^2 C_k \int \mathrm{d}^4x\sqrt{g}\,g^{\mu\nu}g^{\alpha\beta}(\partial_\mu\phi)(\partial_\nu \phi)(\partial_\alpha \phi)(\partial_\beta \phi)\,,
	\label{GlobalSym2}
\end{equation}
where $C_k$ is the corresponding coupling constant. This interaction is compatible with \eqref{eq:shift} and therefore a natural candidate for a matter self-interaction present at a gravity-matter fixed point. Introducing the dimensionless coupling $c_k \equiv C_k k^4$ and substituting \eqref{GlobalSym2} into the FRG \eqref{OFRG2} one concludes that the beta function associated with $c_k$ must be a polynomial of order two in this coupling
\begin{equation}
	\beta_{c} = a_0 + a_1 \, c_k + a_2 \, c^2_k \, . 
	\label{GlobalSym3}
\end{equation}
 The coefficients $a_0$, $a_1$, and $a_2$ collect the dependence of the beta function on the other couplings (as, e.g., Newton's coupling). From \eqref{GlobalSym3}, it then follows that a Gaussian fixed point where $c^* = 0$ requires $a_0 = 0$. Generically, the contribution of $\EuScript{I}_{\rm kin}$ will lead to $a_0 \not = 0$ though: the diagrams constructed from its vertices depend on  the gravitational couplings that contribute to the graviton propagator and the scalar wave-function renormalization only and are independent of $c$. As a result, they induce a non-zero $a_0$ in \eqref{GlobalSym3}. Thus, $c^* = 0$ does not correspond to a fixed point when gravitational couplings have a non-trivial fixed point value. This results in a shift of the Gaussian fixed point (GFP) to a non-vanishing value indicating that one obtains a shifted-Gaussian fixed point (sGFP). Besides the sGFP the theory might have additional interacting fixed points which we refer to as NGFPs. This situation is illustrated in the gray solid and orange dashed curves shown in Fig.\ \ref{sgv3}.
This example illustrates that interactions in the matter sector which respect the symmetries of the kinetic term will generically have non-zero values for their couplings at the NGFP. Thus, the relation \eqref{eq:matterfixedpoints} is modified to 
\be\label{eq:matterfixedpoints2}
\text{NGFP}^{\rm ASQGM} = \underbrace{\text{NGFP}}_{\text{shift-symmetric interactions}} \otimes \underbrace{\text{GFP}^{\rm matter}}_{\text{non-shift symmetric interactions}}   \, .
\ee
Notably, any interaction in the gravitational sector (which by definition is independent of $\phi$) trivially satisfies \eqref{eq:shift} and is thus part of the subspace of shift-symmetric interactions. This structure reconciles the results on Gaussian and non-Gaussian matter couplings available in the literature.

The structure \eqref{GlobalSym3} entails an immanent danger. If the couplings entering $a_0$ conspire to make this coefficient sufficiently large, the shift of the GFP may be so violent that it ceases to exist on the real line. This is illustrated by the blue dotted curve in Fig.~\ref{sgv3}. This suggests that there is an upper bound on the effective interaction strength in the gravitational sector above which an ASQGM-system can not be realized. This has been called the weak-gravity bound \cite{Eichhorn:2016esv,Eichhorn:2017eht}.

Owed to this destabilization-mechanism, investigations of the fixed point structure on the space of shift-symmetry preserving interactions is crucial for establishing the existence of suitable sGFPs and NGFPs underlying the ASQGM-systems. Notably, the class of shift-symmetric interactions contains a wide class of momentum-dependent matter self-interactions like the example given in eq.\ \eqref{GlobalSym2} but also non-minimal gravity-matter interactions. Examples of the latter type include
\be\label{intex1}
\EuScript{I}_3 \sim \int\mathrm{d}^4x\sqrt{g}\,R\,X\,,\qquad \EuScript{I}_4 \sim \int\mathrm{d}^4x\sqrt{g}\,R^{\mu\nu}\,X_{\mu\nu}\, .
\ee
Here we introduced the short-hand notation
\be\label{eq:Xdef}
X_{\alpha\beta} \equiv (\p_\alpha\phi)(\p_\beta \phi) \, , \qquad X \equiv g^{\alpha\beta} X_{\alpha\beta} \, ,
\ee
as the manifestly shift-invariant tensors obtained from the first derivatives of $\phi$.
The interactions \eqref{intex1} are readily generalized by introducing suitable functions, as, e.g., 
\begin{eqnarray}
\EuScript{I} &\sim& \int\mathrm{d}^4x\sqrt{g}\,f(R)\,X \,,\quad \nonumber \\
\EuScript{I} &\sim& \int\mathrm{d}^4x\sqrt{g}\,F(\mathbf{Ric})^{\mu\nu}\,X_{\mu\nu}\,, \\
\EuScript{I} &\sim& \int\mathrm{d}^4x\sqrt{g}\,G(R,\mathbf{Ric},\mathbf{Riem})^{\mu\nu}\,K(X_{\alpha\beta})_{\mu\nu}\,, \nonumber
\label{GlobalSym4}
\end{eqnarray}
where $f(R)$, $F(\mathbf{Ric})^{\mu\nu}$, $G(R,\mathbf{Ric},\mathbf{Riem})^{\mu\nu}$, and $K(X_{\alpha\beta})_{\mu\nu}$ are generic functions. In principle, all these structures should be considered when investigating the non-trivial sector of gravity-matter systems. In this work, we take a key step in this direction: for the first time, the interactions \eqref{GlobalSym2} and \eqref{intex1} are considered simultaneously, taking all backreactions into account. The distinction between such interactions is achieved thanks to the fact that we leave the background metric arbitrary while employing a background approximation.

\section{RG-invariant subspaces of scalar-tensor theories}
\label{subsect.3.2}
It is well-known that in coarse-graining procedures, many symmetries can be broken either by quantum fluctuations (anomalies) or artifically by the regularization process. Given the relevance of shift symmetry in the case of scalar field theories coupled to gravity, one might wonder if such a symmetry resists the regularization procedure and the inclusion of quantum effects, i.e., if the effective average action preserves the symmetry along the RG flow. 
In this section, we answer this question by deriving the functional constraint due to shift symmetry that must be satifisfied by $\Gamma_k$. The resulting identity provides a non-perturbative proof that the subspace spanned by shift-symmetric interaction monomials is closed under RG flows.
\begin{figure}[t]
	\begin{center}
		\includegraphics[scale=0.8]{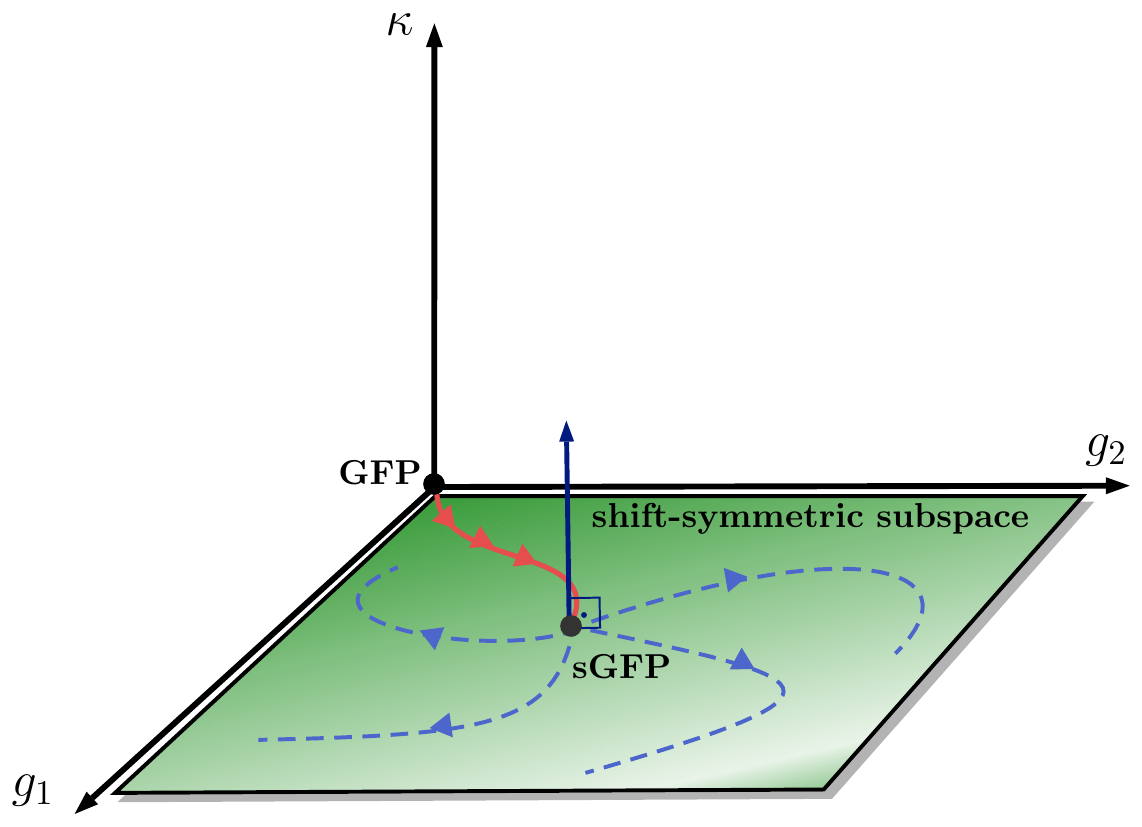}
	\end{center}
	\caption{\label{Fig:shiftsymmetricspace} Illustration of the foliation structure established for the space of matter interactions based on eq.\ \eqref{sswi8}: the subspace of shift-invariant interactions containing the blue-dotted RG trajectories is closed under RG transformations. At the same time non-trivial interactions in the gravitational sector turn on interactions for the shift-symmetric gravity-matter interactions, leading to a sGFP in this sector (red line). The blue arrow attached to the sGFP indicates that generating non-shift-symmetric interactions from an ASQGM-fixed point requires at least one UV-attractive direction pointing outside of the subspace spanned by the shift-symmetric interactions.}
\end{figure}

Let us start by considering a shift-symmetric bare action $S[\varphi]$ which is invariant under \eqref{eq:shift}, i.e., $S[\varphi] = S[\varphi+c]$. The Euclidean path integral is defined as\footnote{Besides the scalar field $\varphi$, $S$ and the path-integral measure include additional fields comprising the spacetime metric $g_{\mu\nu}$ and additional matter fields. Since these do not affect the derivation, we keep these dependencies implicit in order to keep the notation light.}
\begin{equation}
\EuScript{Z}[J] = \int\mathcal{D}\varphi\,\mathrm{e}^{-S[\varphi]+\int_x \,J(x)\varphi(x)}\,,
\label{sswi1}
\end{equation}
where $\int_x \equiv \int {\rm d}^dx \sqrt{g}$. The scale-dependent path integral $\EuScript{Z}_k [J]$ is obtained by adding the regulator \eqref{OFRG1} to $S[\varphi]$,
\begin{equation}
\EuScript{Z}_k [J] = \int\mathcal{D}\varphi\,\mathrm{e}^{-\Sigma[\varphi,J]}\,,
\label{sswi2}
\end{equation}
with
\be\label{eq:defSigma}
\Sigma[\varphi,J] \equiv S[\varphi]+\Delta S_k [\varphi]-\int_x\,J(x)\varphi(x) \, . 
\ee
Since $\varphi$ is a dummy variable, we can replace it by $\varphi^\prime$ in \eqref{sswi2}. Let us assume that $\varphi^\prime = \varphi + \varepsilon$, with $\varepsilon$ being an infinitesimal constant associated with an infinitesimal global shift of $\varphi$. Assuming that the functional measure is invariant under such a transformation, i.e., $\mathcal{D}\varphi^\prime = \mathcal{D}\varphi$, one finds
\begin{eqnarray}
\EuScript{Z} _k [J] &=& \int \mathcal{D}\varphi^\prime \,\mathrm{e}^{-S[\varphi^\prime]-\Delta S_k [\varphi^\prime]+\int_x\,J(x)\varphi^\prime (x)} \nonumber \\ &=& \int \mathcal{D}\varphi\,\mathrm{e}^{-S[\varphi+\varepsilon]-\Delta S_k [\varphi+\varepsilon]+\int_x\,J(x)(\varphi(x)+\varepsilon)}\nonumber\\
&\simeq& \int \mathcal{D}\varphi\,\mathrm{e}^{-\Sigma [\varphi,J]+\varepsilon\int_x \frac{\delta\Sigma}{\delta\varphi}} \\
 &\simeq& \int \mathcal{D}\varphi\,\Bigg(1+\varepsilon \int_x\,\frac{\delta\Sigma [\varphi,J]}{\delta\varphi (x)}\Bigg)\mathrm{e}^{-\Sigma [\varphi,J]}\nonumber\\
&=&\EuScript{Z}_k [J]+\varepsilon\,\int\mathcal{D}\varphi\int_x\,\frac{\delta\Sigma [\varphi,J]}{\delta\varphi (x)}\mathrm{e}^{-\Sigma [\varphi,J]}\, , 
\nonumber
\label{sswi3}
\end{eqnarray}
where we expanded up to linear order in $\epsilon$.
Thus, one concludes that
\be\label{eq:finalshift1}
\int\mathcal{D}\varphi\int_x\,\frac{\delta\Sigma [\varphi,J]}{\delta\varphi (x)} \, \mathrm{e}^{-\Sigma [\varphi,J]} = 0 \, . 
\ee
Substituting the definition \eqref{eq:defSigma} one then has
\begin{eqnarray}
\int_x\frac{\delta\Sigma [\varphi,J]}{\delta\varphi (x)} & = & \int_x\Bigg(\frac{\delta S}{\delta\varphi (x)}+\frac{\delta\Delta S_k}{\delta\varphi (x)}-J(x)\Bigg) \nonumber \\ &=& \int_x\Bigg(\frac{\delta\Delta S_k}{\delta\varphi (x)}-J(x)\Bigg)\,,
\label{sswi4}
\end{eqnarray}
where shift-symmetry implies that the term containing the action $S$ is a surface term which vanishes once integrated over spacetime. Hence, eq.\ \eqref{eq:finalshift1} simplifies to
\begin{eqnarray}
\int\mathcal{D}\varphi\int_x\Bigg(\frac{\delta\Delta S_k}{\delta\varphi (x)}-J(x)\Bigg)\mathrm{e}^{-\Sigma [\varphi,J]} = 0\,.
\label{sswi5}
\end{eqnarray}
Since $\Delta S_k$ is quadratic in $\varphi$, we have $\frac{\delta\Delta S_k[\varphi]}{\delta\varphi (x)} = \mathcal{R}_k\,\varphi (x)$. Substituting this expression into the path integral then gives
\begin{eqnarray}
\int\mathcal{D}\varphi\int_x\,\mathcal{R}_k\,\varphi (x)\,\mathrm{e}^{-\Sigma [\varphi,J]} & = & \EuScript{Z}_k [J]\int_x\,\mathcal{R}_k\langle \varphi (x)\rangle_J\nonumber\\
&=&  \EuScript{Z}_k [J] \int_x\,\mathcal{R}_k\,\phi (x) \\ & = & \EuScript{Z}_k [J] \int_x\,\frac{\delta\Delta S_k [\phi]}{\delta\phi (x)}\,.
\label{sswi6}
\end{eqnarray}
Furthermore, the source $J(x)$ can be expressed in terms of the effective average action, exploiting the quantum equation of motion \cite{Reuter:2019byg} 
\begin{equation}
 \frac{\delta\Gamma_k [\phi]}{\delta\phi (x)}+\frac{\delta \Delta S_k[\phi]}{\delta\phi (x)} = J(x)\,.
\label{sswi7}
\end{equation}
Using eqs.\ \eqref{sswi7} and \eqref{sswi6} in \eqref{sswi5} leads to\,,
\begin{equation}
\int_x\,\frac{\delta\Gamma_k [\phi]}{\delta\phi (x)} = 0\,.
\label{sswi8}
\end{equation}
Thus, $\Gamma_k[\phi]$ inherits the invariance under shift-symmetry from the bare action. Notably, this result holds for any regulator, provided that it is quadratic in the field. Taking the derivative of \eqref{sswi8} with respect to the RG time $t$, one furthermore finds that shift symmetry is preserved along the RG flow, \emph{if the symmetry is preserved by the initial conditions}, see Fig\, \ref{Fig:shiftsymmetricspace}. Thus, the space of shift-symmetric actions is closed under RG flows in the sense that the flow does not generate shift-symmetry violating terms dynamically. This has the profound consequence that a scalar potential cannot be generated from the shift-symmetric interactions \emph{unless the symmetry is broken by leaving the NGFP along a UV-attractive direction}.

\section{RG flows of shift-symmetric scalar-tensor theories}
\label{sect.4setup}
In the last section we provided a non-perturbative proof that the space of shift-symmetric scalar-tensor interactions is closed under the RG flow. Moreover, we argued in Sect.\ \ref{Sect.2.2} that understanding the fixed point structure of the couplings associated with these terms is critical for assessing the feasibility of realizing an ASQGM-system. Guided by these insights, we now proceed by studying the RG flow in this sector within a rather advanced approximation of $\Gamma_k$. The main results of this section are the beta functions \eqref{eq:beta-grav}, \eqref{eq:beta-matter}, \eqref{eq:beta-matterII}, and \eqref{eq:anomimplicit} encoding the running and fixed point structure obtained within this computation. Our calculation is performed for four-dimensional Euclidean spacetime and within the background approximation of the FRG.
\subsection{Projecting the RG flow}
\label{ssect.4.1}
Schematically, the effective average action for a scalar-tensor theory has the structure
\be\label{eq:ans-grav-gf}
\Gamma^{\rm scalar-tensor}_k= \Gamma^{\rm grav}_k[g]+ \Gamma_k^{\rm matter}[\phi,g]+\Gamma^{\rm gf}_k + \Gamma_k^{\rm ghost} \,.
\ee
Here $\Gamma_k^{\rm grav}[g]$ captures the gravitational dynamics, $\Gamma_k^{\rm matter}[\phi,g]$ contains the matter sector, and $\Gamma^{\rm gf}_k$ and $\Gamma_k^{\rm ghost}$ are the gauge-fixing and ghost terms complementing the construction. When constructing $\Gamma^{\rm grav}_k[g]$, we assume that the dynamics of the metric degrees of freedom is sufficiently well-approximated by the Einstein-Hilbert truncation,
\be\label{eq:ans-grav}
\Gamma^{\rm grav}_k[g] \simeq \frac{1}{16\pi G_k} \int \mathrm{d}^4x \sqrt{g} \left[2 \Lambda_k -R\right] ,  
\ee
which includes a running Newton coupling $G_k$ and cosmological constant $\Lambda_k$. Our emphasis in the matter sector is on shift-symmetric interactions. Explicitly, we include interactions up to four derivatives
\be\label{eq:ans-matter}
\Gamma_k^{\rm matter}[\phi,g] \simeq \int \mathrm{d}^4x \sqrt{g} \left[ \frac{Z_k}{2} X
+ Z_k^2 \, C_k \, X^2 + Z_k \tilde{C}_k R^{\mu\nu} X_{\mu\nu} + Z_k \, D_k \, R \, X
\right] \, , 
\ee
where $X_{\mu\nu} \equiv (\p_\mu \phi)(\p_\nu \phi)$ and $X \equiv g^{\mu\nu} X_{\mu\nu}$ are introduced in eq.\ \eqref{eq:Xdef}. This ansatz comprises a scale-dependent wave-function renormalization $Z_k$, the coupling associated with the scalar self-interaction $C_k$  (mass dimension $-4$), and couplings of the scalar to the spacetime curvature $\tilde{C}_k$ and $D_k$ (mass-dimension $-2$). Thus the non-minimal terms contained in \eqref{eq:ans-grav} correspond to power-counting irrelevant operators. In order to have a complete basis of four-derivative terms, one should also include the scalar bilinear with second derivatives of the scalar, $\sim\,(D^{2} \phi)^2$, which would essentially correspond to promoting the wave function renormalization $Z_k \mapsto Z_k(-D^2)$ to a momentum-dependent form factor. The inclusion of this factor will be left to future work.

Our gauge-fixing procedure utilizes the background field method \cite{Percacci:2017fkn,Reuter:2019byg,Abbott:1981ke}, building on the decomposition
\be
g_{\mu\nu} = \gb_{\mu\nu} + h_{\mu\nu}\,, \qquad {\rm and} \qquad \phi = \bar{\phi} + \varphi\,.
\ee
Here $\gb_{\mu\nu}$ and $\bar{\phi}$ are arbitrary background fields and $h_{\mu\nu}$ and $\varphi$ are the fluctuations about those backgrounds, which are not necessarily small. From now on, indices are lowered and raised by the background metric and its inverse. In the following, we adopt the harmonic gauge
\be\label{eq:gf}
\Gamma_k^{\rm gf}[h;\gb] = \, \frac{1}{32 \pi G_k} \int \mathrm{d}^4x \sqrt{\gb} \gb^{\mu\nu} F_\mu F_\nu \, , \quad {\rm with} \quad F_\mu = \Db^\nu h_{\mu\nu} - \frac{1}{2} \Db_\mu h \,. 
\ee
Here $\Db_\mu$ is the covariant derivative constructed from the background metric $\gb_{\mu\nu}$ and $h\equiv\gb^{\mu\nu}h_{\mu\nu}$. For $h_{\mu\nu} = 0$, which suffices for a background computation, the Faddeev-Popov action associated with \eqref{eq:gf} is
\be
S^{\rm ghost}[\bar{C},C,h=0;\gb] = -\sqrt{2}\int \mathrm{d}^4x \sqrt{\gb}\,\bar{C}_\mu\, \Big(\delta^{\mu}_\nu \Db^2 + \bar{R}^{\mu}_{\phantom{\alpha}\nu}\Big)C^\nu\,.
\label{ghostaction}
\ee
Here we neglect the wave function renormalization associated with the ghost fields $\bar{C}_\mu,C^\nu$. 

The setup is completed by specifying the regulator $\cR_k$ appearing in the FRG. Following the classification \cite{Codello:2008vh}, we implement a regularization of Type I, utilizing the replacement rule
\be\label{Rregdef}
\Delta \mapsto P_k(\Delta) \equiv \Delta + R_k(\Delta)
\ee
constructed for the background Laplacian $\Delta \equiv - \gb^{\mu\nu} \Db_\mu \Db_\nu$. The dimensionful profile function, defined in eq.\eqref{OFRG1}, is taken to be of Litim-type \cite{Litim:2000ci}
\be\label{litimreg}
R_k(z) = (k^2 - z)\theta(k^2-z) \, . 
\ee
This completes the setup of our computation and constitutes the starting point for solving \eqref{OFRG2} in the given approximation.
\subsection{Evaluating the flow equation}
\label{EvalFlowEq}
We aim to derive the beta functions determining the $k$-dependence of the (dimensionless counterparts of the) couplings $\{G_k, \Lambda_k,C_k, \tilde{C}_k, D_k \}$ as well as the scalar anomalous dimension $\eta_s \equiv - k \p_k \ln Z_k$. This can be achieved by projecting the FRG onto the ansatz \eqref{eq:ans-grav-gf}. The main task here is the evaluation of the trace appearing on the right-hand side of the flow equation \eqref{OFRG2}.
 This can be achieved by means of the universal RG machine developed in \cite{Benedetti:2010nr} and recently upgraded in \cite{Knorr:2021slg}. The core idea is to use the choices of gauge, background, and regulator together with suitable algebraic manipulations to cast the operator that we want to trace into a form where standard heat-kernel techniques can be applied.

As a first step, one computes the Hessian arising from \eqref{eq:ans-grav-gf} in a fixed but arbitrary background
\begin{equation}
\left[\Gamma^{(2)}_k\right]_{IJ} = \frac{1}{\sqrt{\bar{g}}}\frac{1}{\sqrt{\bar{g}}}\frac{\delta^2\Gamma_k}{\delta \Phi_I \delta \Phi_J}\Big|_{\Phi = 0}\,.
\label{eq:genericHessian}
\end{equation}
Here $\Phi_I = \{h_{\mu\nu}, \varphi, \bar{C}_\mu, C^\nu\}$ is the collection of fluctuation fields. We highlight that, in this schematic notation, we are assuming that suitable signs of derivatives with respect to Grassmannian fields are taken into account. Within the background approximation, the elements of the Hessian \eqref{eq:genericHessian} lead to the  general structure,
\begin{equation}
\left[\Gamma^{(2)}_k + \cR_k\right]_{IJ} = \prop_{IJ}(\Delta)+\mathbb{D}_{IJ} (\bar{D}_{\cdot})+\mathbb{V}_{IJ}(\bar{\Phi};\bar{D}_{\cdot})\,.
\label{eq:genericHessian2}
\end{equation}
The function $\prop_{IJ}(\Delta)$ contains the (regularized) kinetic terms of the fluctuation fields and collects all the background Laplace-operator contributions to the Hessian. Typically, this matrix is diagonal in field space. Terms containing uncontracted derivatives $\bar{D}_{\cdot}$ \emph{but no background structures} are collected in $\mathbb{D}_{IJ}$. Finally, all terms containing background structures, such as spacetime curvatures or background matter fields are contained in $\mathbb{V}$. 

The right-hand side of \eqref{OFRG2} contains the inverse of \eqref{eq:genericHessian2}. This inverse can be constructed perturbatively in the background quantities, expanding $\left[\Gamma^{(2)}_k+ \cR_k\right]^{-1}$ and retaining all terms build from the background operators which appear in the expansion basis within our truncation. In order to perform such an inversion and get a suitable expression that enables the application of standard heat-kernel techniques, one needs to get rid of $\mathbb{D}$-type terms. In the present setting, this is achieved by adopting harmonic gauge in which the graviton propagators depend on the Laplacian only and do not contain uncontracted derivatives $\bar{D}_\cdot$. Terms containing the ``background structures'' $\Rb$, $\Rb_{\mu\nu}$, $\bar{C}_{\mu\nu\rho\sigma}$ and $\p_\mu \bar{\phi}$ are grouped into $\mathbb{V}$.

With the reduction of \eqref{eq:genericHessian2} to the desired form, the perturbative inversion is, in a matrix notation, given by
\begin{eqnarray}
\left[\Gamma^{(2)}_k + \cR_k\right]^{-1} &=& \mathbb{P}^{-1}_k \times \left[\mathbb{1}+  \ver\times\mathbb{P}^{-1}_k\right]^{-1} \nonumber\\ &\simeq& \mathbb{P}^{-1}_k - \mathbb{P}^{-1}_k \times \ver \times \mathbb{P}^{-1}_k 
+\mathbb{P}^{-1}_k \times \ver \times  \mathbb{P}^{-1}_k \times \ver \times \mathbb{P}^{-1}_k + \mathcal{O}(\ver^3)\,,
\label{eq:inversionregHessian}
\end{eqnarray}
with $\prop^{-1}_k$ being the matrix of regularized propagators $\gprop$. Since the vertices $\ver$ carry background structures, the expansion can be truncated when it reaches the same power of the background structures as the one present in $\Gamma_k [\Phi = 0; \bar{\Phi}]$. At this stage, the computation has boiled down to evaluating a sum of traces, which can be performed via the off-diagonal heat-kernel techniques of Appendix \ref{App.C}. Once this has been done, the beta functions for the dimensionful couplings can be read off from the coefficients multiplying the various interaction terms contained in the ansatz. Note that no physical meaning should be attributed to these terms at this point. They merely serve as book-keeping devices spanning a (truncated) basis on the space of all possible scalar-tensor theories.

\subsection{Traces, propagators, vertices, and projection rules}
\label{sect.3.1}
We now apply the general algorithm of Sect.\ \ref{EvalFlowEq} to the specific truncation \eqref{eq:ans-grav-gf}. We first note that the components of the regularized kinetic terms, $\prop_{IJ}$, for the graviton fluctuations, scalar fluctuations, and ghosts are
\be\label{eq:Pkres}
\begin{split}
	\left[ \prop_h \right]_{\mu\nu}{}^{\alpha\beta} = & \, \frac{1}{32\pi G_k} \, (\Delta + R_k) \, \left[  \Pi_{\phantom{\text{TL}}\mu\nu}^{\text{TL}\phantom{\mu\nu}\alpha\beta}
	-  \Pi_{\phantom{\text{T}}\mu\nu}^{\text{T}\phantom{\mu\nu}\alpha\beta} \right] \, , \qquad
	\prop_{\varphi}  =  \, Z_k \, (\Delta + R_k) \, , \\
	\left[ \prop_{\bar{C}C} \right]_{\mu}{}^{\nu} = & \, (\Delta + R_k) \, \delta_\mu^\nu \, .
\end{split}
\ee
Here $\Pi^{\text{TL}}$ and $\Pi^{\text{T}}$ are projection operators on the space of symmetric matrices whose definitions are given in Eq.\ \eqref{eq:projectorsdef}. The entries of the propagator matrix $\gprop$ are readily obtained from inverting \eqref{eq:Pkres}
\be\label{eq:Gdef}
\begin{split}
\left[ \gprop^{h} \right]_{\mu\nu}{}^{\alpha\beta} = & \, \frac{32 \pi G_k}{(\Delta + R_k)} \left[ \Pi_{\phantom{\text{TL}}\mu\nu}^{\text{TL}\phantom{\mu\nu}\alpha\beta} - \Pi_{\phantom{\text{T}}\mu\nu}^{\text{T}\phantom{\mu\nu}\alpha\beta} \right] \, , \qquad
\gprop^{\varphi} =  \,  \frac{1}{Z_k \, (\Delta + R_k) } \, , \\
\left[ \gprop^{\bar{C}C} \right]_{\mu}{}^{\nu} = & \, \frac{1}{(\Delta + R_k)}  \delta_\mu^\nu \, . 
\end{split}
\ee
The matrix entries of the regulator $\cR_k$ are readily read off from \eqref{eq:Pkres} and are given by
\be\label{eq:Rdef}
\begin{split}
\left[ \cR_k^h \right]_{\mu\nu}{}^{\alpha\beta} = & \, \frac{1}{32\pi G_k} \,  R_k  \, \left[  \Pi_{\phantom{\text{TL}}\mu\nu}^{\text{TL}\phantom{\mu\nu}\alpha\beta}
-  \Pi_{\phantom{\text{T}}\mu\nu}^{\text{T}\phantom{\mu\nu}\alpha\beta} \right] \, , \qquad
\cR_k^\varphi  =  \, Z_k \, R_k \, , \\
\left[ \cR_k^{\bar{C}C} \right]_{\mu}{}^{\nu} = & \, R_k\, \delta_\mu^\nu \, . 
\end{split}
\ee

In addition to the propagators one also needs the explicit expressions for the interaction vertices $\ver$. These are conveniently generated using the xAct package suite for {\tt Mathematica} \cite{xActwebpage,2007CoPhC.177..640M,Brizuela:2008ra,2008CoPhC.179..597M,2014CoPhC.185.1719N}. Since the resulting expressions are rather bulky, they have been relegated to Appendix \ref{App.B}.

The ansatz \eqref{eq:ans-grav-gf} entails that the information about the scale-dependence of the coupling is encoded in the coefficients multiplying the following interaction monomials
\be\label{eq:basisdef}
\begin{array}{lll}
	\EuScript{I}_0 \equiv \int {\rm d}^4x \sqrt{\gb} \, , \quad & 
	\EuScript{I}_1 \equiv \int {\rm d}^4x \sqrt{\gb}\, \bar{R} \, , \quad  &
	\EuScript{I}_2 \equiv \int {\rm d}^4x \sqrt{\gb}\, \bar{X} \, , \\
	\EuScript{I}_3 \equiv \int {\rm d}^4x \sqrt{\gb}\, \bar{R}\bar{X} \, , \quad &
	\EuScript{I}_4 \equiv \int {\rm d}^4x \sqrt{\gb}\, \bar{R}^{\mu\nu}\bar{X}_{\mu\nu} \, , \quad & 
	\EuScript{I}_5 \equiv \int {\rm d}^4x \sqrt{\gb}\, \bar{X}^2 \, . 
\end{array}
\ee
The next step then constitutes in substituting the expansion \eqref{eq:inversionregHessian} into the trace appearing on the right-hand side of eq.\ \eqref{OFRG2} and identifying the terms which give rise to contributions proportional to at least one of the elements \eqref{eq:basisdef}. Analyzing the structure of the background fields appearing in the vertices listed in Appendix \ref{App.B} shows that the flow on the subspace spanned by our ansatz receives contributions from 18 traces
\be
\frac{1}{2}\mathrm{STr}\left[\left(\Gamma^{(2)}_k + \mathcal{R}_k\right)^{-1}\partial_t\mathcal{R}_k\right] \simeq \sum^{18}_{n=1} T_n \, . 
\ee
The first eight traces contain the graviton regulator and read
\be\label{gravity-traces}
\begin{split}
	T_1 = & \, \frac{1}{2} {\rm Tr}_{(2)} \left[ \gprop^h \cdot \p_t \cR_k^h \right] \, , \\
	T_2 = & \, - \frac{1}{2} {\rm Tr}_{(2)} \left[ \gprop^h \cdot \ver_{hh} \cdot \gprop^h \cdot \p_t \cR_k^h \right] \, , \\
	T_3 = & \, \frac{1}{2} {\rm Tr}_{(2)} \left[ \gprop^h \cdot \ver_{h\varphi} \, \gprop^\varphi \, \ver_{\varphi h} \cdot \gprop^h \cdot \p_t \cR_k^h \right] \, , \\
	T_4 = & \, \frac{1}{2} {\rm Tr}_{(2)} \left[ \gprop^h \cdot \ver_{hh} \cdot \gprop^h \cdot \ver_{h h} \cdot \gprop^h \cdot \p_t \cR_k^h \right] \, , \\
	T_5 = & \, - \frac{1}{2} {\rm Tr}_{(2)} \left[ \gprop^h \cdot \ver_{hh} \cdot \gprop^h \cdot \ver_{h \varphi} \, \gprop^{\varphi} \,  \ver_{\varphi h} \cdot \gprop^h \cdot \p_t \cR_k^h \right] \, , \\
	T_6 = & \, - \frac{1}{2} {\rm Tr}_{(2)} \left[ \gprop^h \cdot \ver_{h \varphi} \, \gprop^{\varphi} \,  \ver_{\varphi h} \cdot \gprop^h \cdot \ver_{hh} \cdot \gprop^h  \cdot \p_t \cR_k^h \right] \, , \\
	T_7 = & \, \frac{1}{2} {\rm Tr}_{(2)} \left[ \gprop^h \cdot \ver_{h\varphi} \, \gprop^{\varphi} \,  \ver_{\varphi h} \cdot \gprop^h \cdot \ver_{h \varphi} \, \gprop^{\varphi} \,  \ver_{\varphi h} \cdot \gprop^h  \cdot \p_t \cR_k^h \right] \, , \\
	T_8 = & \, - \frac{1}{2} {\rm Tr}_{(2)} \left[ \gprop^h \cdot \ver_{h \varphi} \, \gprop^{\varphi} \,  \ver_{\varphi \varphi} \, \gprop^\varphi \,  \ver_{\varphi h} \cdot \gprop^h  \cdot \p_t \cR_k^h \right] \, . \\
\end{split}
\ee
Here the subscript $(2)$ on the trace indicates that the trace is on the space of symmetric matrices and the symbol $\cdot$ marks a contraction of spacetime indices. In addition there are eight traces involving the scalar regulator. Their structure is identical to the one found in the gravitational sector \eqref{gravity-traces} with $h \leftrightarrow \varphi$ interchanged and the trace taken with respect to scalar fluctuations:
\be\label{scalar-traces}
\begin{split}
	T_9 = & \, \frac{1}{2} {\rm Tr}_{(0)} \left[ \gprop^{\varphi} \, \p_t \cR_k^\varphi \right] \, , \\
	T_{10} = & \, - \frac{1}{2} {\rm Tr}_{(0)} \left[ \gprop^{\varphi} \, \ver_{\varphi\varphi} \, \gprop^{\varphi} \, \p_t \cR_k^\varphi \right] \, , \\
	T_{11} = & \, \frac{1}{2} {\rm Tr}_{(0)} \left[ \gprop^{\varphi} \, \ver_{\varphi h} \cdot \gprop^h \cdot \ver_{ h \varphi} \, \gprop^{\varphi} \, \p_t \cR_k^\varphi \right] \, , \\
	T_{12} = & \, \frac{1}{2} {\rm Tr}_{(0)} \left[ \gprop^{\varphi} \, \ver_{\varphi\varphi} \, \gprop^{\varphi} \, \ver_{\varphi\varphi} \, \gprop^{\varphi} \, \p_t \cR_k^\varphi \right] \, , \\
	T_{13} = & \, - \frac{1}{2} {\rm Tr}_{(0)} \left[ \gprop^{\varphi} \, \ver_{\varphi\varphi} \, \gprop^{\varphi} \, \ver_{\varphi h} \cdot \gprop^{h} \cdot  \ver_{h \varphi} \, \gprop^{\varphi} \, \p_t \cR_k^\varphi \right] \, , \\
	T_{14} = & \, - \frac{1}{2} {\rm Tr}_{(0)} \left[ \gprop^{\varphi} \; \ver_{ \varphi h} \cdot \gprop^{h} \cdot  \ver_{h \varphi} \; \gprop^{\varphi} \; \ver_{\varphi\varphi} \; \gprop^{\varphi}  \; \p_t \cR_k^\varphi \right] \, , \\
	T_{15} = & \,  \frac{1}{2} {\rm Tr}_{(0)} \left[ \gprop^{\varphi} \; \ver_{ \varphi h} \cdot \gprop^{h} \cdot  \ver_{ h \varphi} \; \gprop^{\varphi} \; \ver_{\varphi h} \cdot \gprop^{h} \cdot  \ver_{h \varphi} \; \gprop^{\varphi}  \; \p_t \cR_k^\varphi \right] \, , \\
	T_{16} = & \, - \frac{1}{2} {\rm Tr}_{(0)} \left[ \gprop^{\varphi} \; \ver_{\varphi h} \cdot \gprop^{h} \cdot  \ver_{hh} \cdot \gprop^h \cdot  \ver_{h \varphi} \; \gprop^{\varphi}  \; \p_t \cR_k^\varphi \right] \, . \\
\end{split}
\ee
The ghosts contribute
\be\label{ghost-traces}
\begin{split}
	T_{17} = & \, - {\rm Tr}_{(1)} \left[ \gprop^{\bar{C}C} \cdot \p_t \cR_k^{\bar{C}C} \right] \, , \qquad
	T_{18} =  \,  {\rm Tr}_{(1)} \left[ \gprop^{\bar{C}C} \cdot \ver_{\bar{C}C} \cdot \gprop^{\bar{C}C} \cdot \p_t \cR_k^{\bar{C}C} \right] \, ,
\end{split}
\ee
with the trace taken on the space of vectors.

The evaluation of these traces can be simplified significantly by noting that certain background structures do not contribute to the flow projected onto the ansatz spanned by \eqref{eq:basisdef}. Hence, the computation can be simplified by equating the corresponding structures to zero. A careful analysis shows that this applies to
\be\label{eq:projectionrules}
\bar{C}_{\mu\rho\nu\sigma} \sim 0 \, , \qquad  (\bar{D}_\mu \bar{R}_{\alpha\beta}) \sim 0 \, , \quad \mathrm{and} \quad (\bar{D}_\alpha \bar{R}) \sim 0 \, , 
\ee
as well as symmetrized strings of covariant derivatives acting on the background scalar field,
\be\label{eq:projectionrules2}
\bar{D}_{(\mu_1} \cdots \bar{D}_{\mu_n)} \bar{\phi} \sim 0 \, , \quad n \ge 2 \,  . 
\ee
Here the symbol $\sim$ is used to indicate that these terms are not necessarily zero. They merely generate interaction monomials not tracked in our truncation. We stress that the projection rules \eqref{eq:projectionrules} and \eqref{eq:projectionrules2} do not entail a particular choice of background so that it is clear that any result derived by imposing these relations is still manifestly background-independent. While the conditions \eqref{eq:projectionrules} state that any derivative of a curvature tensor will not contribute to interaction monomials contained in our ansatz, a remark about our projection condition for matter fields is in order. The ansatz \eqref{eq:ans-matter} contains first derivatives of the scalar field $\bar{\phi}$ only. Therefore, it is tempting to impose a projection condition which eliminates all terms where two or more covariant derivatives act on $\bar{\phi}$. In a flat background, this would correspond to projecting out contributions which contain more than one power of the  momentum associated with the corresponding scalar field. In a curved spacetime, the commutator of covariant derivatives gives rise to curvature terms which in turn will generate contributions proportional to $\EuScript{I}_3$ and $\EuScript{I}_4$. Imposing that one drops terms containing \emph{symmetrized derivatives}, as indicated by the condition \eqref{eq:projectionrules2}, then reduces to the intuitive picture of dropping terms containing higher powers of the momentum in a flat background while, at the same time, keeping track of the curvature terms generated by anti-symmetric combinations of the derivatives.

Equipped with these prerequisits the traces \eqref{gravity-traces}, \eqref{scalar-traces}, and \eqref{ghost-traces} can be evaluated based on the heat-kernel techniques reviewed in Appendix \ref{App.C}. Since the resulting expressions are rather bulky, we have relegated these intermediate results to Appendix \ref{App.D} where we also add some comments on technical subtleties. This forms the basis for the beta functions given in the next section.

\subsection{Beta functions}
\label{Sect.5}
We are now in a position to write down the beta functions controlling the scale-dependence of the couplings $\{G_k, \Lambda_k, C_k, \tilde{C}_k, D_k\}$ and the scalar wave function renormalization $Z_k$. For this purpose it is convenient to trade the dimensionful couplings with their dimensionless counterparts
\be\label{dimless}
g_k \equiv G_k k^2 \, , \quad \lambda_k \equiv \Lambda_k k^{-2} \, , \quad \tilde{c}_k \equiv \tilde{C}_k k^2 \, , \quad d_k \equiv D_k k^2 \quad c_k \equiv C_k k^4 \, , 
\ee
and introduce the anomalous dimensions $\eta_s$ and $\eta_N$ associated with the wave function renormalization of the scalar field and Newton's coupling as
\be\label{anomdim}
\eta_s \equiv - k \p_k \ln Z_k \, , \qquad \eta_N \equiv k \p_k \ln G_k \, . 
\ee
Denoting the collective set of dimensionless couplings by $u_i \equiv \{g_k,\lambda_k, c_k, d_k, \tilde{c}_k\}$ the running of the couplings is encoded in the beta functions
\be\label{betadef}
k \p_k u_i = \beta_{u_i}(u_j) \, , \qquad \qquad i = 1,\cdots, 5\, .
\ee
The beta function associated with the dimensionless Newton's coupling can be expressed through the anomalous dimension $\eta_N$, $\beta_g = (2 + \eta_N) \, g$.

Matching the coefficients multiplying the monomials \eqref{eq:basisdef} on the left- and right-hand side of the FRG gives a coupled set of equations determining the scale-dependence of the quantities \eqref{dimless}. Following \cite{Reuter:1996cp}, the result is conveniently expressed in terms of the dimensionless threshold functions defined in eqs.\ \eqref{thresholdeval} and \eqref{eq:defq}. The short-hand notation on the upper index in the $q$-functionals appearing in Appendix \ref{App.D} is lifted by counting the number of scalar and graviton propagators in the corresponding trace. Since all dimensionless $q$-functionals are evaluated at argument $-2\lambda$, we omit this information for the sake of readability. Furthermore, it is understood that all couplings depend on the scale $k$, so that the subscripts are omitted as well. This said, the beta functions encoding the scale-dependence of $g_k$ and $\lambda_k$ are
\be\label{eq:beta-grav}
\begin{split}
	\beta_\lambda = & \, (\eta_N - 2) \lambda + \frac{g}{2\pi}\left( 10 \, q^{0,1}_{N,2} + q^{1,0}_{s,2} - 8 \, \Phi^{1,0}_2 \right) \, , \\ 
	\beta_g = & \, (2 + \eta_N) \, g \, .
\end{split}
\ee
The three beta functions associated with the non-minimal interaction terms are
\be\label{eq:beta-matter}
\begin{split}
	\beta_{\tilde{c}} = & \, (2+\eta_s) \ct 
	\\ & \, - \frac{g}{3\pi} \big( q_{N,2}^{1,2} + (9 \ct + 10 d)  q_{N,2}^{0,2} - 6(3 \ct +2 d) q_{N,3}^{1,2} -3 (7 \ct - 10 d) (\ct + 2 d) q_{N,4}^{1,2} \\ & \qquad \qquad -180 \ct (\ct + 2 d)^2 q_{N,6}^{2,2} \big) 
	\\ & \, + \frac{c}{12 \pi^2} q_{s,2}^{2,0} + \frac{c \, \ct}{\pi^2} q_{s,4}^{3,0} + \frac{g}{\pi} \big(4 (\ct+2 d) q_{s,3}^{2,1} - (11 \ct+6 d) \, (\ct+2 d) q^{2,1}_{s,4} \\ & + 120 \,  \ct \,  (\ct+2 d)^2 q^{3,1}_{s,6} \big) \, , \\[1.2ex]
	\beta_{d} = & \, (2+\eta_s) d \\ & \,
	+ \frac{g}{6\pi} \Big( 4 q^{1,2}_{N,2} -18 q^{1,3}_{N,3} + 180 (\ct + 2 d)^2 (\ct + 6 d) q^{2,2}_{N,6} + 3 (\ct + 2 d) (7 \ct + 22 d) q^{1,2}_{N,4} \\ & \quad \quad \quad
	-72 (\ct + 2 d)^2 q^{1,3}_{N,5}  -2 (3 \ct + 14 d) q^{0,2}_{N,2} -18 (3 \ct + 4 d) q^{0,3}_{N,3} + 6 (3 \ct + 10 d) q^{1,2}_{N,3} \\ & \quad \quad \quad -36 (\ct + 2 d) q^{1,3}_{N,4} -18 (\ct + 4 d) q^{2,2}_{N,4}  \Big) \\ & \,
	-\frac{c}{12 \pi ^2} q^{2,0}_{s,2} + \frac{c (2 \ct+9 d)}{\pi ^2} q^{3,0}_{s,4} 
	+ \frac{g}{6 \pi } \big( q^{2,1}_{s,2} -9 q^{2,2}_{s,3} + 12 (\ct + 4 d) q^{2,1}_{s,3}\\ & \qquad -3 (\ct - 6 d) (\ct + 2 d) q^{2,1}_{s,4}   -18 (\ct + 2 d) q^{2,2}_{s,4} -36 (\ct + 2 d)^2 q^{2,2}_{s,5}
	-36 (\ct + 4 d) q^{3,1}_{s,4} \\ & \qquad+ 360 (\ct + 2 d)^2 (\ct + 6 d) q^{3,1}_{s,6} \big) \, , \\[1.2ex]
\end{split}
\ee
\be
\begin{split}\label{eq:beta-matterII}
	\beta_{c} = & \, 2 \, (2 + \eta_s)  c \\
	& \, + \frac{12 g}{\pi} \Big( 
	2 c q^{1,2}_{N,3} -c \, q^{0,2}_{N,2} 
	+ 2 c (\ct + 2 d) q^{1,2}_{N,4}
	-3 c q^{2,2}_{N,4}
	+ 60 c (\ct + 2 d)^2  q^{2,2}_{N,6}
	+ 4 g \pi q^{0,3}_{N,2} \\ & \qquad
	+ 4 \pi  g \left(5 \ct^2+4 \ct d-2 d^2\right) q^{0,3}_{N,4}
	-4 \pi g  q^{1,3}_{N,3}
	-8 \pi  g (\ct + 2 d) q^{1,3}_{N,4}
	+ 88 \pi g (\ct + 2 d)^2  q^{1,3}_{N,5} \\ & \qquad
	+ 40 \pi g (\ct + 2 d)^2 (7 \ct + 18 d) q^{1,3}_{N,6}
	+ 8 \pi g q^{2,3}_{N,4}
	-240 \pi g  (\ct + 2 d)^2 q^{2,3}_{N,6} \\ & \qquad
	+ 7560 \pi g (\ct + 2 d)^4 q^{2,3}_{N,8} 
	-48 \ct g^2 q^{0,3}_{N,3}
	\Big) \\ & \, 
	+ \frac{3}{\pi^2}\Big(
	5 c^2 q^{3,0}_{s,4} 
	+ 8 \pi g  c q^{2,1}_{s,3}
	+ 8 \pi g c (\ct + 2 d) q^{2,1}_{s,4}
	-24 \pi g c  q^{3,1}_{s,4}
	+ 480 \pi g c (\ct + 2 d)^2 q^{3,1}_{s,6} \\ & \qquad
	-8 \pi^2 g^2 q^{2,2}_{s,3} 
	-16 \pi^2 g^2 (\ct + 2 d)   q^{2,2}_{s,4}
	+ 80 \pi^2 g^2 (\ct + 2 d)^2 (7 \ct + 18 d)  q^{2,2}_{s,6} \\ & \qquad
	+ 32 \pi^2 g^2 q^{3,2}_{s,4}
	+ 176 \pi^2 g^2 (\ct + 2 d)^2 q^{2,2}_{s,5} 
	-960 \pi^2 g^2 (\ct + 2 d)^2 q^{3,2}_{s,6} \\ & \qquad
	+ 30240 \pi^2 g^2 (\ct + 2 d)^4 q^{3,2}_{s,8}
	\Big) \, . 
\end{split}
\ee
The system is completed by the expressions for the anomalous dimensions which can be read off from the coefficients multiplying $\EuScript{I}_1$ and $\EuScript{I}_2$: 
\be\label{eq:anomimplicit}
\begin{split}
	\eta_N = & \, \frac{g}{6 \pi} \left(10  \, q_{N,1}^{0,1}  - 36 \, q_{N,2}^{0,2} +     q^{1,0}_{s,1} -  6 (\tilde{c} + 4 d) \, q^{2,0}_{s,3} - 8  \, \Phi^1_1 - 12 \,  \Phi^2_2 \right) \, , \\
	\eta_s = & \frac{1}{2\pi^2} \Big(
	72 \pi g  (\ct + 2 d)^2 (q^{1,2}_{N,5} + q^{2,1}_{s,5})
	-24 \pi g (\ct + d) q^{0,2}_{N,3} 
	- 8 \pi g (q^{1,2}_{N,3} + q^{2,1}_{s,3})
	+ 3 c \, q^{2,0}_{s,3}
	\Big) \, . 
\end{split}
\ee
The explicit form of $\eta_N$ and $\eta_s$ as a function of the couplings is obtained by  solving this linear system of equations. Together with the beta functions \eqref{eq:beta-grav}, \eqref{eq:beta-matter}, and \eqref{eq:beta-matterII}, these results completely determine the scale-dependence of the couplings \eqref{dimless}.

\section{Fixed point structure}
\label{Sect.fps}
We are now in the position to give an elaborate discussion of the fixed point structure supported by the beta functions \eqref{eq:beta-grav}, \eqref{eq:beta-matter}, and \eqref{eq:beta-matterII}. We start by commenting on the general structure of $\beta_c$, $\beta_{\ct}$ and $\beta_d$ in Sect.\ \ref{sect.5.1} before discussing the properties of the NGFPs in Sect.\ \ref{sect.5.2}. We close the investigation by considering the fixed points appearing in ``pure-matter'' truncations in a curved (non-fluctuating) background spacetime in Sect.\ \ref{Sect.5a}. 
\subsection{Structure of the beta functions in the matter sector}
\label{sect.5.1}
We start the investigation by looking at the structure of the matter-beta functions. In the approximation where the scalar anomalous dimension is set to zero, they are given by polynomials in the non-minimal matter couplings. The coefficients depend on the gravitational couplings $\lambda, g$. When the effect of the scalar anomalous dimension is included, the structure of the beta functions changes and they are given by the quotient of two polynomials. In order to exhibit this structure, we pick a ``reference point'' for the gravitational couplings,
\be\label{eq:gravbp}
g_{\rm ref} = 0.66 \, , \qquad \lambda_{\rm ref} = 0.2 \, . 
\ee
These values are suggested by the position of the non-Gaussian gravity-matter fixed point seen at minimal coupling. In a first instance, it is then instructive to consider a one-dimensional projection of the non-minimal matter beta functions, setting the couplings of the other operators to zero by hand. Schematically, $\beta_g$ then takes the form
\be\label{eq:polymatter}
\beta_{g_m} = \frac{p(g_m)}{q(g_m)} \, , \qquad p(g_m) = \sum_{i=0}^{n} p_i(g,\lambda) (g_m)^i \, , \qquad  q(g_m) = \sum_{i=0}^{\tilde{n}} q_i(g,\lambda) (g_m)^i \, . 
\ee
Here we normalized $q_0 = 1$ and $g_m$ denotes the dimensionless version of the single matter coupling in the one-dimensional projection. The polynomial structure of the resulting systems is then compiled in Table \ref{Tab.polyorder}.
\begin{table}[t!]
	\centering
	\begin{tabular}{c|c||c|cc||cc}
	\hline
	 & $\eta_s$ & $p_0(g_{\rm ref}, \lambda_{\rm ref})$ & $n$ & $\tilde{n}$ & $n$ & $\tilde{n}$ \\
	\hline
	\multirow{2}{*}{$\beta_c$} & \XSolid & $\phantom{-}54.0$ & $2$ & $0$   & $2$ & $0$ \\ 
	& \Checkmark & $\phantom{-}54.0$ & $3$ & $1$ & $3$ & $1$ \\
	\hline
	\multirow{2}{*}{$\beta_d$} & \XSolid & $-0.47$ & $3$ & $0$ & $3$ & $0$  \\ 
	& \Checkmark & $-0.48$ & $6$ & $3$  & $6$ & $3$ \\ \hline
	\multirow{2}{*}{$\beta_{\tilde{c}}$} & \XSolid & $-0.13$ & $3$ & $0$  & $3$ & $0$  \\ 
	& \Checkmark & $-0.13$ & $6$ & $3$ & $6$ & $3$ \\ \hline
	\end{tabular}
\caption{\label{Tab.polyorder} Polynomial structure of the matter beta-functions evaluated at the base-point \eqref{eq:gravbp} (and $\eta_N = -2$) obtained from the one-dimensional projection (middle block) and including the full backreaction of all non-minimal gravity matter couplings (last two columns).}
\end{table}
There are several observations linked to this data:
\begin{itemize}
	\item[1)] All polynomials have a non-zero value $p_0(g_{\rm ref},\lambda_{\rm ref})$. Thus setting the  matter coupling to zero does not correspond to a fixed point of the corresponding beta function. This is the mechanism creating the ``shifted Gaussian fixed points'': the presence of non-zero gravitational couplings lead to non-zero interactions in the matter sector.
	\item[2)] The actual value of $p_0$ is insensitive to the inclusion of the scalar anomalous dimension.
	\item[3)] If $\eta_s$ is set to zero $n$ may be even or odd. Beta functions where $n$ is odd automatically have at least one real zero, extending the gravitational fixed point to a fixed point with a non-zero matter coupling. If $n$ is even, there are three possible scenarios: the polynomial may have two, one, or zero real roots. In the first case, one observes a splitting of the gravitational fixed point into two gravity-matter fixed points. The critical exponents generated by the matter-coupling (essentially encoding the slope of the beta function in the direction of the matter coupling) come with opposite signs. Hence, it is expected that the two gravity-matter fixed points generated by the inclusion of the non-minimal matter coupling come with different predictive power.
	\item[4)] The inclusion of the scalar anomalous dimension changes the degree $n$ of $p(g_m)$ \emph{without altering the index} $n - \tilde{n}$. In this way, the system can generate additional zero's of the beta functions. In the following, we adopt the selection criterion that a viable NGFP must be present in both systems with $\eta_s$ included and set to zero. In this viewpoint the scalar anomalous dimension has the status of a correction to the fixed point properties (refining the approximation) instead of being the key ingredient generating the fixed point.\footnote{Note that this does not rule out that there can also be viable RG fixed points of the second category where the anomalous dimension is essential for creating the fixed point. The fact that fluctuations in the matter sector are essential for these fixed points puts their study beyond the scope of the present work.}
	\item[5)] The inclusion of the backreaction from the other non-minimal matter couplings does not affect the polynomial structure exhibited in eq.\ \eqref{eq:polymatter}. In this case, the coefficients depend on both the gravitational couplings $\{\lambda,g\}$ as well as the other non-minimal gravity matter couplings though.
	\item[6)] The only beta function with an even index $n-\tilde{n}$ in its associated matter coupling is $\beta_c$. Thus it is this direction in theory space which is ``dangerous'' in the sense that the corresponding beta function may not exhibit real zeros as indicated in the left panel of Fig.\ \ref{sgv3}.
\end{itemize}

\subsection{Fixed points of the gravity-matter system}
\label{sect.5.2}
Following up on the general structure of the beta functions in the matter sector, we now determine the fixed point structure of the system. Besides the full system, comprising the five scale-dependent couplings $\{\lambda,g,c,\ct,d\}$, it is also instructive to consider the six subtruncations where either two or one of the couplings in the matter sector is set to zero and the corresponding beta function is dropped from the set of equations. It is then convenient to label the resulting truncations by the matter couplings retained in the approximation. In addition, the fixed point structure can be analyzed including the scalar anomalous dimension or in the approximation where $\eta_s = 0$ by hand. The inclusion of $\eta_s$ will then be indicated by adding the symbol ``$*$'' to the truncation label. In this nomenclature, the full system is then denoted by $c\ct d^*$. 

\begin{table}[p!]
	\begin{center}
		\begin{tabular}{lccccccccccc}
			& $g_*$ & $\lambda_*$ & $c_*$ & $\tilde{c}_*$ & $d_*$ & $\eta_s$ & $\theta_1$ & $\theta_2$ & $\theta_3$ & $\theta_4$  & $\theta_5$ \\ \hline 
			mc & $0.66$ & $0.21$ & $-$ & $-$ & $-$ & $-$ & \multicolumn{2}{c}{$1.60 \pm 3.25i$} & $-$ & $-$ & $-$  \\ \cline{2-12}
			mc$^*$ & $0.66$ & $0.21$ & $-$ & $-$ & $-$ & $-0.77$ & \multicolumn{2}{c}{$1.60 \pm 3.28i$} & $-$ & $-$ & $-$
			\\ \hline \hline 
			$c$ & $-$ & $-$ & $-$ & $-$ & $-$ & $-$ & \multicolumn{2}{c}{$-$} & $-$  & $-$ & $-$ 
			\\ \cline{2-12} 
		$c^*$	& $-$ & $-$ & $-$ & $-$ & $-$ & $-$ & \multicolumn{2}{c}{$-$} & $-$  & $-$ & $-$  
			\\ \hline \hline 
			$\ct$ & $0.66$ & $0.21$ & $-$ & $\phantom{-}0.08$ & $-$ & $-$ & \multicolumn{2}{c}{$1.60 \pm 3.25i$} & $-$ & $-1.79$   & $-$ 
			\\  \cline{2-12}
			\multirow{3}{*}{$\ct^*$} & $0.65$ & $0.21$ & $-$ & $\phantom{-}0.17$ & $-$ & $-1.03$ & \multicolumn{2}{c}{$1.60 \pm 3.29i$} &  $-$ & $-0.56$   & $-$ \\
			& $0.65$ & $0.21$ & $-$ & $\phantom{-}0.75$ & $-$ & $-1.74$ & \multicolumn{2}{c}{$1.60 \pm 3.28i$} &  $-$ & $\phantom{-}0.40$   & $-$ \\
			& $0.64$ & $0.20$ & $-$ & $\phantom{-}2.20$ & $-$ & $-2.42$ & \multicolumn{2}{c}{$1.65 \pm 3.17i$} &  $-$ & $-1.33$   & $-$
			\\ \hline \hline
			\multirow{3}{*}{$d$} & $0.66$ & $0.21$ & $-$ & $-$ & $-1.24$ & $-$ & \multicolumn{2}{c}{$1.74 \pm 3.47i$} & $-$    & $-$ & $-3.90$ 
			\\ 
			& $0.66$ & $0.21$ & $-$ & $-$ & $-0.22$ & $-$ & \multicolumn{2}{c}{$1.59 \pm 3.33i$} & $-$    & $-$ & $\phantom{-}2.22$ 
			\\ 
			& $0.66$ & $0.19$ & $-$ & $-$ & $\phantom{-}1.20$ & $-$ & \multicolumn{2}{c}{$1.75 \pm 2.96i$} & $-$    & $-$ & $-4.47$ 
			\\ \cline{2-12}
			\multirow{3}{*}{$d^*$} & $0.67$ & $0.21$ & $-$ & $-$ & $-0.74$ & $\phantom{-}1.13$ & \multicolumn{2}{c}{$1.66 \pm 3.33i$} & $-$    & $-$ & $-3.20$ 
			\\ 
			& $0.66$ & $0.21$ & $-$ & $-$ & $-0.19$ & $-0.43$ & \multicolumn{2}{c}{$1.59 \pm 3.33i$} & $-$    & $-$ & $\phantom{-}2.32$ 
			\\ 
			& $0.66$ & $0.19$ & $-$ & $-$ & $\phantom{-}1.31$ & $-0.46$ & \multicolumn{2}{c}{$1.74 \pm 2.94i$} & $-$    & $-$ & $-7.26$ 
			\\ \hline \hline 
			$c\tilde{c}$ &  $-$ & $-$ & $-$ & $-$ & $-$ & $-$ & \multicolumn{2}{c}{$-$} & $-$  & $-$ & $-$ 
			\\ \cline{2-12} 
			$c\tilde{c}^*$ & $-$ & $-$ & $-$ & $-$ & $-$ & $-$ & \multicolumn{2}{c}{$-$} & $-$  & $-$ & $-$
			\\ \hline \hline
			$cd$ &  $0.66$ & $0.21$ & $-80.8$ & $-$ & $-1.98$ & $-$ & \multicolumn{2}{c}{$2.12 \pm 3.44i$} & $-1.71$  & $-$ & $-18.3$ 
			\\ \cline{2-12} 
			$cd^*$ & $0.67$  & $0.21$ & $-16.7$ & $-$ & $-0.93$ & $1.51$ & \multicolumn{2}{c}{$1.58\pm3.48i$} & $-0.25$  & $-$  & $-2.19$
			\\ \hline \hline
			\multirow{3}{*}{$\tilde{c}d$} &  $0.66$ & $0.21$ & $-$ & $\phantom{-}0.33$ & $-1.23$ & $-$ & \multicolumn{2}{c}{$1.67 \pm 3.49i$} & $-$  & \multicolumn{2}{c}{$-2.49 \pm 0.68i$} \\ 
			&  $0.66$ & $0.21$ & $-$ & $\phantom{-}0.05$ & $-0.28$ & $-$ & \multicolumn{2}{c}{$1.59 \pm 3.35i$} & $-$  & $-1.86$  & $\phantom{-}2.17$ \\ 
			&  $0.66$ & $0.19$ & $-$ & $\phantom{-}0.59$ & $\phantom{-}1.15$ & $-$ & \multicolumn{2}{c}{$1.74 \pm 2.95i$} & $-$  & $-3.13$ & $-5.91$
			\\ \cline{2-12}   
			\multirow{3}{*}{$\tilde{c}d^*$} & $0.67$  & $0.21$ & $-$ & $\phantom{-}0.08$  & $-0.74$ & $\phantom{-}0.94$ & \multicolumn{2}{c}{$1.65\pm3.34i$} & $-$  & $-2.62$  & $-2.96$ \\
			& $0.66$  & $0.21$ & $-$ & $\phantom{-}0.07$  & $-0.25$ & $-0.42$ & \multicolumn{2}{c}{$1.58\pm3.34i$} & $-$  & $-1.40$  & $\phantom{-}2.25$ \\
			& $0.66$  & $0.19$ & $-$ & $\phantom{-}0.71$  & $\phantom{-}1.20$ & $-0.44$ & \multicolumn{2}{c}{$1.74\pm2.92i$} & $-$  & $-2.88$  & $-9.43$
			\\ \hline \hline
			\multirow{1}{*}{$c\ct d$} 
			& $0.66$ & $0.21$ & $-29.0$ & $\phantom{-}0.63$ & $-1.69$ & $-$ & \multicolumn{2}{c}{$1.76 \pm 3.57i$} & \multicolumn{2}{c}{$-1.88 \pm 1.28i$} & $-9.69$
			\\ \cline{2-12} 
			\multirow{2}{*}{$c\ct d^*$}  & $0.67$ & $0.21$ & $-16.6$  & $\phantom{-}0.14$  & $-0.96$ & $\phantom{-}1.27$ & \multicolumn{2}{c}{$1.70 \pm 3.38i$} &  \multicolumn{2}{c}{$-4.54 \pm 2.69i$} & $-3.00$ \\
			& $0.64$ & $0.24$ & $-42.2$  & $\phantom{-}5.65$  & $-4.23$ & $-1.75$ & \multicolumn{2}{c}{$1.34 \pm 4.65i$} &  \multicolumn{2}{c}{$-7.52 \pm 1.80i$} & $\phantom{-}2.20$
			\\ \hline \hline
		\end{tabular}
	\end{center}
	\caption{\label{tab.fpdata} We summarize the fixed point structure entailed by our beta functions in various subsystems. Coordinates and critical exponents are given up to two decimal digits. The first two lines refer to a minimally coupled scalar field (mc), providing the benchmark for the truncations including the shift-symmetric interactions in the matter sector. The latter are labeled by the couplings retained in the corresponding subsystem. E.g., the full system tracking the couplings $c, \ct$, and $d$ together with the scalar anomalous dimension $\eta_s$ is referred to as $c\ct d^*$, etc.}
\end{table}
\clearpage
The fixed points identified in the various (sub-)systems are listed in Table \ref{tab.fpdata}. This data constitutes the main result of this section. Notably, the fixed point structure of the systems $c$ and $\ct$ have been studied in \cite{deBrito:2021pyi} and \cite{Eichhorn:2017sok}, respectively. In both cases our computation confirms the results motivated by the hybrid computation reported in these works.

Given that the complexity of the beta functions increases significantly once an additional coupling is included, we apply two different search strategies for identifying their roots. For the simple systems $c, \ct, d$ and $c^*, \ct^*, d^*$ the set of fixed points can be obtained via the {\tt NSolve}-command of {\tt Mathematica}. In these cases, it is possible to verify analytically that the numerical algorithm identified all zeros of the set of beta functions. This strategy underlies the data given in the first four blocks of Table \ref{tab.fpdata}, reporting the fixed point structure at minimal coupling (mc) and upon retaining one non-minimal gravity-matter couplings.

The fixed points listed for the more complex truncations are found in a multistep procedure. First, the beta functions of the matter sector are simplified by substituting the approximate fixed point values of the gravitational couplings given in \eqref{eq:gravbp}. The complete list of zeros of the simplified matter sector is then obtained from {\tt NSolve}, which results in approximate values for $c_*, \ct_*, d_*$. Combined with the base point for $\lambda$ and $g$, this data serves as initial condition for finding the exact position of the root through {\tt Mathematica}'s {\tt FindRoot} algorithm. This strategy has the advantage that it converges rapidly to the exact position of the fixed point. At the same time it allows to determine the number of NGFPs rooted in the same base point in the gravitational sector in a reliable way.  

\begin{figure}[t!]
	\begin{center}
		\includegraphics[scale=0.8]{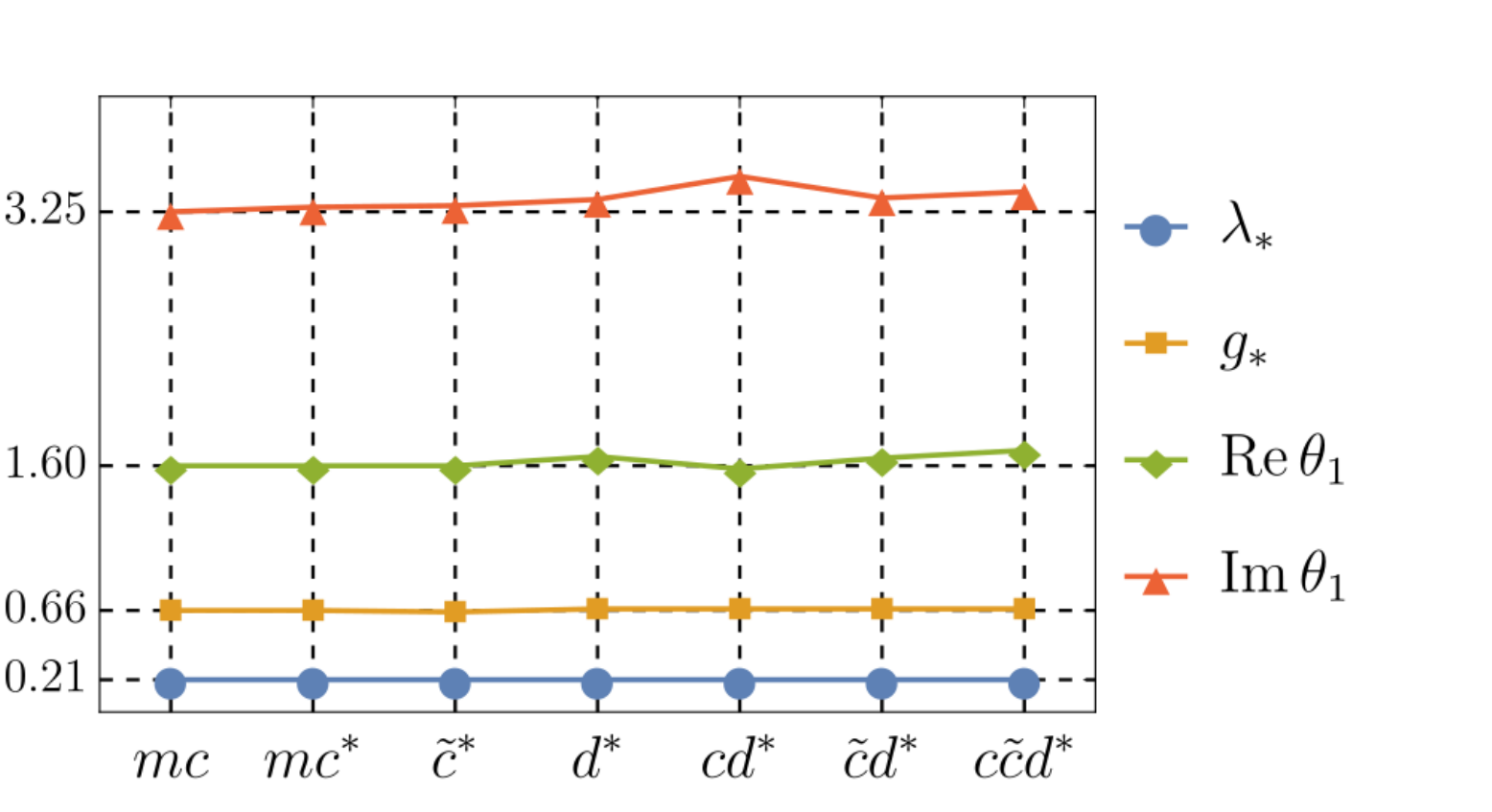}
	\end{center}
	\caption{\label{Fig:plotSTAS} Illustration of the characteristic features linked to the gravitational sector of the NGFPs for various subsystems. The data points are based on Table \ref{tab.fpdata}. Notably, the position and critical exponents are rather insensitive to the truncation employed in the matter sector.}
\end{figure}

At this point the following technical remark is in order. Our analysis also revealed that the interactions in the matter-sector may also generate zeros of the beta functions which are not visible at minimal couplings and their properties in the gravitational sector differ substantially from the ones shown in Fig.\ \ref{Fig:plotSTAS}. While the existence of this new class of solutions is intriguing, the complexity of the coupled system of beta functions makes the systematic inventarization of all potential fixed point candidates technically rather involved. Thus this task is left for a future investigation.

Table \ref{tab.fpdata} allows to draw the following conclusions: \\

\noindent
\emph{Stability of the Gravitational Sector} \\
All fixed points identified in our work essentially share the same properties in the gravitational sector: the position $\{\lambda_*,g_* \}$ as well as the critical exponents $\theta_1, \theta_2$ are almost unaffected by the feedback from the non-minimal interaction terms. This feature is highlighted in Fig.\ \ref{Fig:plotSTAS} which compares the characteristics of the gravitational sector in various truncations.\footnote{When a given truncation gives multiple NGFPs, we select the one closest to the one observed at minimal coupling. Since the values for Re($\theta_1$) and Re($\theta_2$) among the different candidates differ by less than 20\% the feature exhibited in Fig.\ \ref{Fig:plotSTAS} is shared by all candidates.} \\
%

\noindent
\emph{Truncations with a single non-minimal coupling} \\
The NGFP seen at minimal coupling (mc) persists in the $\ct$ and $d$-truncations while it is absent in the $c$-truncation. These features are readily understood from the properties of the beta functions summarized in Table \ref{Tab.polyorder}. $\beta_{\ct}$ and $\beta_d$ are cubic in the respective couplings. Thus they always give rise to at least one real root and the existence of a NGFP is guaranteed. The cubic nature of the matter beta functions can also entail the possibility that the NGFP visible at minimal coupling splits into a triplet of NGFPs which are distinguished by their fixed point coordinate $\ct$ or $d$. These triplets appear in the $\ct^*$, $d$, and $d^*$-truncations. Their characteristic feature is an alternating sign for the critical exponent appearing in the matter sector. This can be traced back to the fact that the derivatives of a cubic polynomial at its three real roots must have alternating signs.

In contrast to $\beta_{\ct}$ and $\beta_d$, $\beta_c$ is quadratic in $c$. The fact that we do not find a suitable NGFP in the $c$-truncation then indicates that the gravitational interactions shift $\beta_c$ upward so that the resulting quadratic polynomial does not admit real roots. This situation is depicted by the top-curve shown in Fig.\ \ref{sgv3}. In Ref.\  \cite{deBrito:2021pyi}, this situation has been paraphrased as the truncation is incompatible with the weak gravity bound. \\

\noindent
\emph{The stabilizing effect of the $RX$-coupling} \\
The $c$-truncation suggests that the gravity-scalar system considered in this work does not constitute an ASQGM-model, since the matter self-interaction $c$ does not develop a suitable fixed point. The $c\ct$ and $c\ct^*$-truncations seem to corroborate this assessment. This analysis does not take the full backreaction of the matter self-interactions into account. Table \ref{tab.fpdata} shows that the coupling $d$ plays a crucial role in stabilizing the system within the weak gravity bound: all truncations including this coupling possess (at least) one NGFP suitable for realizing Asymptotic Safety. Already in the $cd$-truncation this stabilization effect is sufficiently strong to recreate the fixed point. \\

\noindent
\emph{The full system $c\ct d$} \\
The most elaborate truncation considered in the present work, the $c\ct d$ and $c\ct d^\ast$-truncations described in the last block of Table \ref{tab.fpdata}, exhibit a NGFP suitable for realizing asymptotic safety.\footnote{The $c\ct d^*$ truncation gives rise to a second NGFP created by including the scalar anomalous dimension. This fixed point is listed in the second line for completeness. The present discussion focuses on the fixed point in the first line which, by virtue of its stability coefficient in the matter sector, may be identified as the counterpart of the NGFP listed for the the $c\ct d$-truncation.} Its gravitational sector, comprising $\{\lambda_*, g_*, \theta_1, \theta_2\}$ is strikingly similar to the one associated with the NGFP visible at minimal coupling. The critical exponents associated with the non-minimal matter interactions, $\theta_3, \theta_4, \theta_5$, are all negative indicating that the non-minimal gravity-matter interactions are UV-irrelevant. Hence the power-counting irrelevant operators $\EuScript{I}_3, \EuScript{I}_4, \EuScript{I}_5$ introduced in eq.\ \eqref{eq:basisdef}, remain irrelevant upon including quantum corrections from the gravitational and matter sectors. This suggests that the values of $c, \ct$, and $d$ in the effective action actually constitute predictions from Asymptotic Safety which may be benchmarked against structural requirements on the low-energy effective action \cite{Alberte:2020jsk}. \\

In summary, our analysis revealed that the theory space associated with gravity supplemented by a single scalar field possesses a NGFP suitable for asymptotic safety. In the gravitational sector, the properties of this fixed point are very similar to the ones encountered at minimal coupling. The detailed analysis reported in Table \ref{tab.fpdata} reveals that the interplay of the couplings associated with the non-minimal interactions play a critical role for establishing the presence of this fixed point. 

\subsection{Scalar systems in a flat background}
\label{Sect.5a}
Based on the general beta functions in the matter sector \eqref{eq:beta-matter} and \eqref{eq:beta-matterII}, it is also interesting to study the fixed point structure of the ``pure-matter''-system where all contributions involving graviton propagators are switched off. This corresponds to analyzing the RG flow of the scalar matter system in a non-fluctuating background spacetime which can either be flat or curved. For the flat background, the invariants $\EuScript{I}_3$ and $\EuScript{I}_4$ vanish. The corresponding truncation (flat) contains the scalar-kinetic term $\EuScript{I}_2$ and the momentum-dependent scalar self-interactions $\EuScript{I}_5$ only. Thus we are left with the scalar wave-function renormalization and the coupling $c$. Additionally, the curved background truncation (curved) retains the interations $\EuScript{I}_3$ and $\EuScript{I}_4$, so that the set of couplings in this truncation is given by $\{c, \ct, d\}$.
\begin{table}[t!]
	\begin{center}
		\begin{tabular}{clccccccc}
			& & $c_*$ & $\tilde{c}_*$ & $d_*$ & $\eta_s$  & $\theta_3$ & $\theta_4$  & $\theta_5$ \\ \hline
			\multirow{4}{*}{flat} & GFP &  $0$ & $-$ & $-$ & $-$ & $-4$ & $-$ & $-$ \\
			& NGFP &  $-63.2$ & $-$ & $-$ & $-$ & $\phantom{-}4$ & $-$ & $-$ \\ \cline{2-9}
			& GFP &  $0$ & $-$ & $-$ & $0$ & $-4$ & $-$ & $-$ \\
			& NGFP & $-31.9$ & $-$ & $-$ & $-0.90$ & $4.40$ & $-$ & $-$ 
			\\ \hline \hline 
			\multirow{4}{*}{curved}  & GFP &  $0$ & $0$ & $0$ & $-$ & $-4$ & $-2$ & $-2$ \\
			& NGFP & $-63.2$ & $0.15$ & $0.46$ & $-$ & $4$ & $0.4$ & $-1.73$ 
			\\ \cline{2-9}
			 & GFP &  $0$ & $0$ & $0$ & $0$ & $-4$ & $-2$ & $-2$ \\
			& NGFP & $-31.9$ & $0.16$ & $0.49$ & $-0.90$ & $4.40$ & $0.22$ & $-0.95$ \\ \hline \hline
		\end{tabular}
	\end{center}
	\caption{\label{tab.fpdataflat} We summarize the fixed point structure of the matter sector in the absence of gravity, setting $g = 0, \lambda = 0$. The results on a flat Euclidean background (flat) and on a fixed curved background (curved) are given in the first and second line, respectively. In the flat case, the quadratic nature of $\beta_c$ gives rise to a non-trivial fixed point solution (NGFP). This fixed point extends to the curved space truncation retaining all three matter couplings. }
\end{table}
Technically, the beta functions associated with the flat and curved truncations are obtained from \eqref{eq:beta-matter} and \eqref{eq:beta-matterII} by applying
\be\label{eq:gravity-off}
q_{N,q}^{p_1,p_2} \rightarrow 0 \; , \; \;   \forall \{p_1, p_2\} \qquad \text{and} \qquad q_{s,q}^{p_1,p_2} \rightarrow 0  \; , \; \; \forall p_2 \not = 0 \, . 
\ee
The first substitution eliminates the contributions from all traces including the regulator $\cR_k$ in the gravitational sector while the second identity removes all contributions including at least one gravitational propagator. 

The fixed point structure arising from this setting is summarized in Table \ref{tab.fpdataflat}. The result can be understood as follows. In the flat truncation with $\eta_s = 0$, $\beta_c$ is quadratic in $c$ with vanishing constant coefficient
\be
\left. \beta^{\text{flat}}_c \right|_{\eta_s = 0} = 4 c + \frac{5}{8\pi^2} c^2 \, . 
\ee
Hence we obtain a GFP ($c_* = 0$) and a NGFP ($c_* = - \frac{32\pi^2}{5}$). Since the slope of the polynomial has different signs at these two points, the GFP is UV-repulsive while the NGFP is an UV-attractor. This structure persists upon including the scalar anomalous dimension (second block in the flat rows) and when transiting to the curved space truncation (bottom lines). We note that the inclusion of the scalar-anomalous dimension $\eta_s$ gives rise to a second NGFP solution coming with a large positive $\eta_s$. This zero of the beta function is not included in the table. In addition to that, this model was also explored in \cite{deBrito:2021pyi} as a particular case of their analysis and a NGFP with similar qualitative features - and reasonable quantitative agreement with our findings was reported.

Note that the occurrence of the NGFP in the flat approximation does not conflict with the fixed point structure for Galileon theories recently reported in \cite{Steinwachs:2021jft} which identified a GFP only. As argued in Sect.\ 3, only the inclusion of specific couplings convert a Gaussian into a non-Gaussian matter fixed point. The coupling $c$, which is responsible for generating the NGFP in our case is not included in the study \cite{Steinwachs:2021jft}, which provides a natural explanation for the diverging results. 

\section{Conclusions and outlook}
\label{Sect.6}
The formulation of a consistent and predictive theory of quantum gravity including the observed (and, perhaps, yet to be discovered) matter degrees of freedom is a daunting task. Nevertheless, there are good prospects of achieving this goal in the form of an asymptotically safe quantum gravity-matter (ASQGM) theory. This work has taken important steps towards this goal by studying the renormalization group (RG) flow on the theory space spanned by gravity supplemented by a real scalar field. The kinetic term of the scalar field exhibits a global shift-symmetry being invariant under the transformation $\phi \mapsto \phi + c$ with $c$ being an arbitrary real number. In Sect.\ \ref{subsect.3.2} we demonstrated that this symmetry plays a key role in understanding the structure of the theory space: action functionals invariant under this shift-symmetry span a subspace on which the flow equation for the effective average action closes, i.e., if the initial conditions for the flow respect the shift-symmetry the symmetry is preserved along the RG flow.

This finding has profound consequences for understanding the structure of potential RG fixed points of the theory which are the key element in the asymptotic safety construction. By definition, the gravitational sector of an asymptotically safe gravity-matter system must exhibit non-vanishing gravitational interactions. These interactions are transmitted into the matter sector, where the fixed point then develops non-minimal interactions as well. The ``non-Gaussian'' subspace of the matter sector is spanned by interactions compatible with shift-symmetry. While it is consistent to set all couplings associated with shift-symmetry breaking interaction terms to zero, the non-Gaussian matter interactions play a crucial role in determining whether a theory exhibits asymptotic safety.

In this work, we have tackled the question: May scalar-tensor theories have a high-energy completion through the asymptotic safety mechanism? Based on the general discussion above, the focus of the investigation was on analyzing the effect of shift-symmetric non-minimal interactions. By the mechanisms exhibited in Fig. \ref{sgv3}, these come with the imminent danger of destroying a fixed point by shifting the zeros of the matter beta functions into the complex plane. Concretely, we considered a truncation of the theory space comprising the Einstein-Hilbert action supplemented with a quartic scalar self-interaction as well as the two non-minimal terms coupling scalar-bilinears to the Ricci scalar and the Ricci tensor (cf.\ eq.\ \eqref{eq:ans-matter}). This setting produced two key insights. Firstly, the full truncation possesses an interacting RG fixed point suitable for Asymptotic Safety. Secondly, the interplay of the non-minimal couplings is crucial for establishing this result: in particular the non-minimal coupling involving the Ricci scalar plays a crucial role in ensuring that there is a fixed point with real values of the coupling constants. Thus our work provides strong indications that scalar-tensor theories can indeed have a high-energy completion based on the asymptotic safety mechanism.

On the technical side, our computation followed the spirit of \cite{Groh:2011vn} and evaluated the beta functions using the background field method in an unspecified background. Thus our results are manifestly background independent. While extending the present truncation by including further operators is challenging, the computational toolbox developed in this work does not face any conceptual issue to deal with such refinements on a conceptual level. Our results on the fixed point structure then suggest that such a refinement should primarily focus on including further shift-symmetric non-minimal gravity matter interactions, as these may challenge the existence of the NGFPs found at minimal coupling.  In the best case, such an extended analysis will reveal the onset of stability and convergence of the fixed point structure similarly to the $f(R)$-type computations carried out in the context of pure gravity \cite{Codello:2007bd,Machado:2007ea,Falls:2018ylp,Kluth:2020bdv}.

From a phenomenological viewpoint, a theory of quantum gravity coupled to a single scalar field is still far beyond from a realistic description of our Universe and its matter content. Nonetheless, this setting provides an interesting arena for phenomenological applications, e.g., in the context of cosmology and the physics of black holes. In particular, the so-called Horndeski theories \cite{Horndeski:1974wa,Nicolis:2008in,Deffayet:2009wt}, recently reviewed in \cite{Kobayashi:2019hrl}, have been explored in quite some detail with regards to cosmology \cite{Qiu:2011cy,Kobayashi:2011nu}, black holes \cite{Sotiriou:2011dz,Sotiriou:2013qea,Herdeiro:2015waa}, gravitational waves \cite{Ezquiaga:2017ekz}, and the consistency of the low-energy effective field theory \cite{Alberte:2020jsk}. Our findings support the viability of a scalar-tensor theory as being a fundamental quantum field theory. Thus, a clear direction to pursue is the investigation of quantum-gravity imprints in scalar-tensor theories. In particular, the foliation of the theory space suggests that the phenomenology related to shift-symmetric scalar-tensor theories may have a distinguished role from a fundamental perspective. The fact that the fixed points identified in our work provided substantial predictive power, fixing the value of either two or all three gravity-matter couplings in \eqref{eq:ans-matter} in terms of Asymptotic Safety, makes this setting particularly interesting for confronting the theory with phenomenological implications.
\section*{Acknowledgments}
The authors are grateful to Jan Ambj\o rn, Gustavo P. de Brito, and Astrid Eichhorn for useful discussions. ADP acknowledges CNPq under the grant PQ-2 (309781/2019-1), FAPERJ under the “Jovem Cientista do Nosso Estado” program (E26/202.800/2019), and NWO under the VENI Grant (VI.Veni.192.109) for financial support. JW acknowledges the China Scholarship Council (CSC) for financial support. The work of C.L. is funded by Becas Chile, ANID-PCHA/2020-72210073.
\appendix
\section{Definitions of tensor structures}
\label{App.A}
Throughout the work (anti-)symmetrization is with unit strength, i.e., $X_{(\alpha\beta)} = \frac{1}{2} (X_{\alpha\beta}+X_{\beta\alpha})$, etc.
We define $\Delta \equiv - \gb^{\mu\nu} \Db_\mu \Db_\nu$. The units on the space of symmetric matrices (2),  vectors (1), and scalars (0) are
\be\label{eq:units}
\unit_{{(2)} \, \alpha\beta}{}^{\mu\nu} \equiv \frac{1}{2}\left( \delta_\alpha^\mu \delta_\beta^\nu + \delta_\beta^\mu \delta_\alpha^\nu  \right) \, , 
\qquad 
\unit_{(1) \, \mu}{}^{\nu} \equiv \delta_\mu^\nu \, , \qquad
\unit_{(0)} \equiv 1 \, . 
\ee

On the space of symmetric 2-tensors it is convenient to introduce the following tensors 
\be\label{eq:projectorsdef}
\Pi_{\phantom{\text{TL}}\alpha\beta}^{\text{TL}\phantom{\alpha\beta}\mu\nu} \equiv 
\frac{1}{2}\left( \delta_\alpha^\mu \delta_\beta^\nu + \delta_\beta^\mu \delta_\alpha^\nu  \right) - \frac{1}{4} \gb_{\alpha\beta} \gb^{\mu\nu}
\, , \quad
\Pi_{\phantom{\text{T}}\alpha\beta}^{\text{T}\phantom{\alpha\beta}\mu\nu} \equiv \frac{1}{4} \gb_{\alpha\beta} \gb^{\mu\nu} \, . 
\ee
$\Pi^{\text{TL}}$ and $ \Pi^{\text{T}}$ project a symmetric matrix onto its traceless (TL) and tracepart (T). The projectors are orthogonal in the sense that
\be\label{eq:projectorprop}
\left[ \Pi^{\text{TL}} \right] \cdot \left[ \Pi^{\text{TL}} \right] = \left[ \Pi^{\text{TL}} \right] \, , \quad
\left[ \Pi^{\text{T}} \right] \cdot \left[ \Pi^{\text{T}} \right] = \left[ \Pi^{\text{T}} \right] \, , \quad
\left[ \Pi^{\text{TL}} \right] \cdot \left[ \Pi^{\text{T}} \right] = 
\left[ \Pi^{\text{T}} \right] \cdot \left[ \Pi^{\text{LT}} \right] = 0  \, .
\ee

In order to write the interaction vertices in a compact form we use the short-hand notation introduced in \eqref{eq:Xdef}. Moreover, we use the standard colon-notation for covariant derivatives $D_\alpha \phi \equiv \phi_{;\alpha}$.

\section{Vertices entering the computation of the beta functions}
\label{App.B}
The vertices arising from the second variation of \eqref{eq:ans-matter} with respect to the fluctuation fields are quite bulky. Therefore, we collect the corresponding expressions entering into the evaluation of \eqref{gravity-traces} and \eqref{scalar-traces} in this appendix. At this stage it is understood that all fields and spacetime curvatures are background quantities, i.e., we omit the bar on the metric for the sake of readability. The complete list of vertices is then given in Table \ref{tab:vertex}.
\begin{table}[t!]
	\centering
	\renewcommand{\arraystretch}{1.3}
	\begin{tabular}{ll}
		\hline
		vertex	&	matrix element 	\\
		\hline\hline
		$\left[{\ver}_{hh} \right]_{\alpha\beta}{}^{\mu\nu}$ & 
		$\frac{1}{32\pi G_k} \left[ \frac{2}{3} R \,  \Pi_{\phantom{\text{TL}}\alpha\beta}^{\text{TL}\phantom{\alpha\beta}\mu\nu}
		-2C_{(\alpha\phantom{\mu}\beta)}^{\phantom{(\alpha}\mu\phantom{\beta)}\nu} \right]$
		\\[1.4ex]
		$\left[{\ver}_{\bar{C}C} \right]^{\mu}{}_{\nu}$ & $- R^\mu{}_\nu$ 
		\\[1.4ex]
		\hline\hline
		$\left[{\ver}_{hh} \right]_{\alpha\beta}{}^{\mu\nu}$ & 
		$- \frac{1}{4}  X   \left[\Pi_{\phantom{\text{TL}}\alpha\beta}^{\text{TL}\phantom{\alpha\beta}\mu\nu} - \Pi^{\rm T}{}_{\alpha\beta}{}^{\mu\nu} \right] 
		- \frac{1}{4} \left( g_{\alpha\beta} \, X^{\mu\nu} + X_{\alpha\beta} \, g^{\mu\nu} \right) + \delta_{(\alpha}^{(\mu} X_{\beta)}^{\nu)} $ \\[1.4ex] 
		$\left[{\ver}_{h\phi} \right]_{\alpha\beta}$ & $\frac{1}{2} g_{\alpha\beta} \phi^{; \lambda} D_\lambda - \phi_{;(\alpha} D_{\beta)}$ 
		\\[1.4ex]
		$\left[{\ver}_{\phi h} \right]^{\mu\nu}$ & $- \frac{1}{2} g^{\mu\nu}  D_\lambda   \phi^{; \lambda} + D^{(\mu} \phi^{; \nu)} $ 
		\\[1.4ex] \hline \hline
		$\left[{\ver}_{hh} \right]_{\alpha\beta}{}^{\mu\nu}$ & $-\frac{1}{2} \, X^2 \, \left[\Pi_{\phantom{\text{TL}}\alpha\beta}^{\text{TL}\phantom{\alpha\beta}\mu\nu} - \Pi^{\rm T}{}_{\alpha\beta}{}^{\mu\nu} \right]  + 4 \delta^{(\mu}_{(\alpha} X^{\nu)}_{\beta)} X - \left[ g_{\alpha\beta} X^{\mu\nu} + X_{\alpha\beta} g^{\mu\nu} \right] X$\\
		$$ & $+2 X_{\alpha\beta} X^{\mu\nu}$ \\[1.4ex]
		$\left[{\ver}_{h\phi} \right]_{\alpha\beta}$ & $2 g_{\alpha\beta} X  \phi^{; \lambda} D_\lambda - 4 X \phi_{;(\alpha} D_{\beta)} - 4 X_{\alpha\beta} \phi^{; \lambda} D_\lambda$
		\\[1.4ex]
		$\left[{\ver}_{\phi h} \right]^{\mu\nu}$ & $-2 D_\lambda  \phi^{;\lambda} X g^{\mu\nu} + 4 D^{(\mu} X \phi^{; \nu)} + 4 D_\lambda \phi^{; \lambda} X^{\mu\nu}$ \\[1.4ex]
		$\left[{\ver}_{\phi \phi} \right]$ & $-4 D_\lambda X D^\lambda - 8 D_\lambda X^{\lambda\sigma} D_\sigma$ 
		\\[1.4ex] \hline \hline
		$\left[{\ver}_{hh} \right]_{\alpha\beta}{}^{\mu\nu}$ & \eqref{eq:RicX-0order} + \eqref{eq:RicX-1aorder} + \eqref{eq:RicX-1order} \\[1.4ex]
		$\left[{\ver}_{h \phi} \right]_{\alpha\beta}$ & $g_{\alpha\beta} R^{\lambda\sigma} \phi_{;\lambda} D_\sigma - 2 R^\lambda{}_{(\alpha} \phi_{;\lambda} D_{\beta)} - 2 R^\lambda{}_{(\alpha} \phi_{;\beta)} D_\lambda $ \\
		& $- g_{\alpha\beta} D^\lambda D^\sigma \phi_{;\lambda} D_\sigma + D^\lambda D_{(\alpha} \phi_{;\lambda} D_{\beta)} + D_\lambda D_{(\alpha} \phi_{;\beta)} D^\lambda - D^2 \phi_{;(\alpha} D_{\beta)}$ \\[1.4ex]
		$\left[{\ver}_{\phi h} \right]^{\mu\nu}$ & $- D_\lambda \, R^{\lambda\sigma} \, \phi_{;\sigma} \, g^{\mu\nu} + 2 D^{(\mu} R^{\nu)\lambda} \, \phi_{;\lambda} + 2 D^\lambda R_\lambda{}^{(\mu} \, \phi^{;\nu)}$ \\
		& $+ D^\lambda \, \phi_{;\sigma} D_\lambda D^\sigma g^{\mu\nu} - D^{(\mu} \, \phi_{;\lambda} \, D^{\nu)} D^\lambda - D^\lambda \, \phi^{;(\mu} \, D^{\nu)} D_\lambda + D^{(\mu} \phi^{;\nu)} \, D^2$ \\[1.4ex]
		$\left[{\ver}_{\phi \phi} \right]$ & $-2 D_\lambda R^{\lambda\sigma} D_\sigma$ 
		\\[1.4ex] \hline \hline
		$\left[{\ver}_{hh} \right]_{\alpha\beta}{}^{\mu\nu}$ &  \eqref{eq:RX-0order} + \eqref{eq:RX-1aorder} + \eqref{eq:RX-1order} \\[1.4ex]
		$\left[{\ver}_{h \phi} \right]_{\alpha\beta}$ & 
		$-2 R_{\alpha\beta} \phi_{; \lambda} D^\lambda + g_{\alpha\beta} R\; \phi_{; \lambda} D^\lambda - 2 R \phi_{;(\alpha} D_{\beta)} + 2 D_{(\alpha} D_{\beta)} \phi_{; \lambda} D^\lambda$ \\
		$$ & $-2 g_{\alpha\beta} D^2 \phi_{;\lambda} D^\lambda$ 
		\\[1.4ex]
		$\left[{\ver}_{\phi h} \right]^{\mu\nu}$ & $2 D_\lambda \phi^{; \lambda} R^{\mu\nu} - D_\lambda \phi^{;\lambda} g^{\mu\nu} R + 2 D^{(\mu} \phi^{;\nu)} R - 2 D_\lambda \phi^{;\lambda} D^{(\mu} D^{\nu)}$\\
		$$ & $+2 D_\lambda \phi^{;\lambda} D^2 g^{\mu\nu}$ \\[1.4ex]
		$\left[{\ver}_{\phi \phi} \right]$ & $-2 D_\lambda \, R \, D^\lambda $ 
		\\[1.4ex] \hline \hline
	\end{tabular}
	\caption{\label{tab:vertex} Vertices encoding the interactions between the background and fluctuation fields. Derivatives not denoted by the colon-notation act onto everything to their right. The vertices derived from the Einstein-Hilbert action, the scalar kinetic term, the scalar self-interaction, scalars coupled to the Ricci tensor, and the scalars coupled to the Ricci scalar are provided in the first, second, third, fourth, and fifth block, respectively. The couplings in the matter sector are set to $1$. The definitions of the tensorial quantities are provided in App.\ \ref{App.A}.}
\end{table}

Notably, the expressions $\left[{\ver}_{hh} \right]_{\alpha\beta}{}^{\mu\nu}$ arising from the variation of the curvature-scalar interaction terms are rather involved. For computational convenience all expressions are ordered according to the powers of the spacetime curvature contained, i.e.,
\be\label{eq:curvexp}
\left[{\ver}_{hh} \right]_{\alpha\beta}{}^{\mu\nu} = \sum_{n=0} \, \left[{\ver}_{hh}^{(n)} \right]_{\alpha\beta}{}^{\mu\nu}
\ee
The matter-sector \eqref{eq:ans-matter} gives vertices with $n=0,1$ only.

The second variation of $\int \mathrm{d}^4x \sqrt{g} R^{\mu\nu} X_{\mu\nu}$ with respect to the metric fluctuation generates the following vertex structure. At zeroth order in the curvature one obtains
\be\label{eq:RicX-0order}
\begin{split}
\left[{\ver}_{hh}^{(0)} \right]_{\alpha\beta}{}^{\mu\nu} \sim & \, 
\frac{1}{2} \left( X^{\lambda\sigma} D_\lambda D_\sigma \right) \left[\Pi_{\phantom{\text{TL}}\alpha\beta}^{\text{TL}\phantom{\alpha\beta}\mu\nu} - 3 \, \Pi^{\rm T}{}_{\alpha\beta}{}^{\mu\nu} \right] \\
& \, + \frac{1}{2}\left( g^{\mu\nu} X^\lambda{}_{(\alpha} \left(D_\lambda D_{\beta)} + D_{\beta)} D_\lambda \right) + g_{\alpha\beta} X_\lambda{}^{(\mu} \left( D^\lambda D^{\nu)} + D^{\nu)} D^\lambda \right) \right) \\
& \, + X^{(\mu}_{(\alpha} \, \delta^{\nu)}_{\beta)} D^2  - \frac{1}{2} X^{(\mu}_{(\alpha} \, \left( D^{\nu)} D_{\beta)} + D_{\beta)} D^{\nu)} \right) - \frac{1}{2} \delta^{(\mu}_{(\alpha} X^{\nu)}_\sigma \left(D^\sigma D_{\beta)} + D_{\beta)} D^\sigma \right) \\
& \,- \frac{1}{2} \delta^{(\mu}_{(\alpha} X^\sigma_{\beta)} \left(D_\sigma D^{\nu)} + D^{\nu)} D_\sigma \right) \,  .
\end{split}
\ee
Here $\sim$ denotes that the vertex has been simplified applying the relations \eqref{eq:projectionrules}. Symmetrization of the covariant derivatives acting on the fluctuation fields gives curvature terms related to commutators
\be\label{eq:RicX-1aorder}
\begin{split}
	\left[{\ver}_{hh}^{(1a)} \right]_{\alpha\beta}{}^{\mu\nu} \sim & \, - \frac{1}{2} X^{(\mu}_{(\alpha} R^{\nu)}_{\beta)} - \frac{1}{4} X^\sigma_{(\alpha} R_{\beta)}{}^{(\mu\nu)}{}_\sigma - \frac{1}{4} X_\sigma^{(\mu} R^{\nu)}{}_{(\alpha\beta)}{}^\sigma \, \\ &
	-\frac{1}{2} \left( \delta^{(\mu}_{(\alpha} X^{\nu)}_\lambda R_{\beta)}{}^\lambda + \delta^{(\mu}_{(\alpha} X_{\beta)}^\lambda R_\lambda{}^{\nu)} 
	 + X_{(\alpha}^\lambda R_\lambda{}^{(\mu\nu)}{}_{\beta)} + X_\lambda^{(\mu} R^\lambda{}_{(\alpha\beta)}{}^{\nu)} \right) \, . 
\end{split}
\ee
In addition, the vertex contains contributions including one power of the spacetime curvature which directly appear at the level of the variation 
\be\label{eq:RicX-1order}
\begin{split}
	\left[{\ver}_{hh}^{(1b)} \right]_{\alpha\beta}{}^{\mu\nu} = & \, - \frac{1}{2} R^{\rho\sigma} X_{\rho\sigma} \, \left[\Pi_{\phantom{\text{TL}}\alpha\beta}^{\text{TL}\phantom{\alpha\beta}\mu\nu} - \Pi^{\rm T}{}_{\alpha\beta}{}^{\mu\nu} \right]  
	- \left( g^{\mu\nu} R^\lambda{}_{(\alpha} X_{\beta)\lambda} + g_{\alpha\beta} R^{\lambda(\mu} X^{\nu)}{}_\lambda \right) \\ & \, 
+ 2 \left( \delta^{(\mu}_{(\alpha} R^{\nu)}{}_\lambda X^\lambda{}_{\beta)} +  \delta^{(\mu}_{(\alpha} R_{\beta)}{}^\lambda X_\lambda{}^{\nu)}\right)  
+ 2 X_{(\alpha}^{(\mu} R_{\beta)}^{\nu)} \, .	
\end{split}
\ee
The coupling constants can be restored by multiplying the vertices with $Z_k \tilde{C}_k$.

The vertex structure originating from the second variation of $\int {\rm d}^4x \sqrt{g} R X$ comes with the following contributions. At zeroth order in the spacetime curvature one obtains
\be\label{eq:RX-0order}
\begin{split}
	\left[{\ver}_{hh}^{(0)} \right]_{\alpha\beta}{}^{\mu\nu} \sim & \,
	\frac{1}{2} \, X \, D^2 \,  \left[\Pi_{\phantom{\text{TL}}\alpha\beta}^{\text{TL}\phantom{\alpha\beta}\mu\nu} - 3 \, \Pi^{\rm T}{}_{\alpha\beta}{}^{\mu\nu} \right] 
	+ \frac{1}{2} X \left( g_{\alpha\beta} D^{(\mu} D^{\nu)} + D_{(\alpha} D_{\beta)} g^{\mu\nu} \right) \\ & \, - \frac{1}{2} X \delta^{(\mu}_{(\alpha} \left( D^{\nu)} D_{\beta)} + D_{\beta)} D^{\nu)} \right) 
	- X_{\alpha\beta} D^{(\mu} D^{\nu)} - X^{\mu\nu} D_{(\alpha} D_{\beta)} \\
& \, + \left( X_{\alpha\beta} g^{\mu\nu} + g_{\alpha\beta} X^{\mu\nu} \right) D^2 \, .
\end{split}
\ee
The commutators arising from the symmetrization of the covariant derivatives entering \eqref{eq:RX-0order} give rise to
\be\label{eq:RX-1aorder}
\begin{split}
	\left[{\ver}_{hh}^{(1a)} \right]_{\alpha\beta}{}^{\mu\nu} \sim & \, 
	- \frac{1}{2} X \left( \delta^{(\mu}_{(\alpha} R^{\nu)}_{\beta)} + R_\alpha{}^{(\mu\nu)}{}_\beta \right) \, . 
\end{split}
\ee
In addition, the vertex contains contributions including one power of the spacetime curvature which directly appear at the level of the variation
\be\label{eq:RX-1order}
\begin{split}
	\left[{\ver}_{hh}^{(1b)} \right]_{\alpha\beta}{}^{\mu\nu} = & \, - \frac{1}{2} R X \, \left[\Pi_{\phantom{\text{TL}}\alpha\beta}^{\text{TL}\phantom{\alpha\beta}\mu\nu} - \Pi^{\rm T}{}_{\alpha\beta}{}^{\mu\nu} \right]  
	+ 2 R \, \delta^{(\mu}_{(\alpha} X^{\nu)}_{\beta)} + 2 X \delta^{(\mu}_{(\alpha} R^{\nu)}_{\beta)} \\ & \,
	- \frac{1}{2} R \left( X_{\alpha\beta} g^{\mu\nu} + X^{\mu\nu} g_{\alpha\beta} \right) - \frac{1}{2} X \left( R_{\alpha\beta} g^{\mu\nu} + R^{\mu\nu} g_{\alpha\beta} \right)
	 + X_{\alpha\beta} R^{\mu\nu} \\
& \, + X^{\mu\nu} R_{\alpha\beta} \, . 	
\end{split}
\ee
The coupling constants can be restored by multiplying the vertices with $Z_k D_k$.
\clearpage
\section{Heat-kernel techniques}
\label{App.C}
In this appendix, we collect the results on the heat-kernel expansion which are used for evaluating the operator traces appearing on the right-hand side of Wetterich's equation \eqref{OFRG2}.
\subsection{Early-time expansion of the heat-kernel}
Standard heat-kernel techniques \cite{Vassilevich:2003xt,Barvinsky:1985an,Codello:2008vh} and their extension to the off-diagonal heat-kernel \cite{Decanini:2005gt,Anselmi:2007eq,Groh:2011dw,Codello:2012kq} provide a powerful tool for computing quantum corrections in a curved spacetime. The central idea is to introduce the heat kernel
\be
H(x,x^\prime;s) \equiv \langle x | \, e^{-s \Delta} \, | x^\prime \rangle \, . 
\ee
as the matrix elements of the operator $e^{-s \Delta}$. The heat kernel solves the heat equation
\be
(\p_s + \Delta_{x^\prime}) H(x,x^\prime;s) = 0\, ,
\ee
subject to the boundary condition
\be
\lim_{s \rightarrow 0} H(x,x^\prime;s) = \delta(x,x^\prime) \, . 
\ee
The trace of $e^{-s \Delta}$ is then encoded in the ``diagonal part'' (coincidence limit) of $H(x,x^\prime;s)$
\be
{\rm Tr}\left[ e^{-s \Delta} \right] = \int {\rm d}^dx \sqrt{g}\,  H(x,x;s) \, . 
\ee

The heat kernel admits various different approximations. For the present purpose, it is convenient to utilize the early-time expansion which by power-counting can be seed to correspond to an expansion of $H(x,x;s)$ in powers of the spacetime curvature and its derivatives
\be\label{eq:early-time}
H(x,x;s) = \frac{1}{(4\pi s)^{d/2}} \sum_{n=0}^\infty {\rm tr}(a_{2n}) \, s^n \, ,
\ee
 where $n$ counts the number of derivatives contained in the expansion coefficient $a_{2n}$ and ${\rm tr}$ is a trace over vector-bundle indices. The present computation requires the expansion coefficients for $n=0,1$ which are given by
 \be
 a_0 = \unit \, , \qquad a_2 = \frac{1}{6} R \, \unit \, .
 \ee
\subsection{Results for the off-diagonal heat-kernel}
Following \cite{Benedetti:2010nr}, we define the expansion of the heat-kernel at non-coincident points
\be\label{eq:offdiag-exp}
H(x,x^\prime;s) = \frac{1}{(4\pi s)^{d/2}} \, e^{- \frac{\sigma(x,x^\prime)}{2s}} \, \sum_{n=0}^\infty \, s^n \, A_{2n}(x,x^\prime) \, .
\ee
Here $\sigma(x,x^\prime)$ is half the squared geodesic distance between $x$ and $x^\prime$ and we introduced the off-diagonal heat-kernel coefficients $ A_{2n}(x,x^\prime)$, $n \in \mathbb{N}$. The coincident limit $x = x^\prime$ of any object will be denoted by an overbar. For $\sigma(x,x^\prime)$ we then have $\overline{\sigma(x,x^\prime)} = 0$ and the only non-trivial coincidence limit appears if we take two covariant derivatives
\be\label{eq:sigmacoincidence}
\overline{\sigma(x,x^\prime)_{;(\mu\nu)}} = g_{\mu\nu} \, . 
\ee
Up to terms of $\cO(R^2)$, which do not contribute to the present computation, the non-trivial coincident limits of the coefficients $A_{2n}$ and their derivatives are
\be\label{eq:Acoincidence}
\overline{A_0(x,x^\prime)} = \unit \, , \qquad \overline{A_0(x,x^\prime)_{;(\mu\nu)}} = \frac{1}{6} R_{\mu\nu} \, \unit \, , \qquad \overline{A_2(x,x^\prime)} = \frac{1}{6} R \, \unit \, , 
\ee
where $\unit$ denotes the unit operator on the vector bundle on which $\Delta$ acts (cf.\ \eqref{eq:units}).

When applying the off-diagonal heat-kernel in the trace evaluation, one frequently encounters covariant derivatives acting on \eqref{eq:offdiag-exp}. The resulting structures motivate defining
\be\label{eq:Hdef}
H_{\alpha_1 \cdots \alpha_{2n}}(x;s) \equiv \overline{D_{(\alpha_1} \cdots D_{\alpha_{2n})} H(x,x^\prime; s)} \, . 
\ee
The tensors $H_{\alpha_1 \cdots \alpha_n}(x;s)$ are symmetric in all indices by construction. Their early-time expansion can be constructed by taking covariant derivatives of \eqref{eq:offdiag-exp} and evaluating the coincidence limit by substituting the expressions \eqref{eq:Acoincidence}. For the present computation it suffices to carry out the early time expansion up to first order in the spacetime curvature and up to $n=2$. The explicit form of the $H$-tensors is \cite{Benedetti:2010nr}
\be
\begin{split}
H_{\alpha\beta} = \frac{1}{(4\pi s)^{d/2}}  & \, \left[- \frac{1}{2s} g_{\alpha\beta} \left( 1 + \frac{1}{6} s R \right) + \frac{1}{6} R_{\alpha\beta} \right] \unit \, , \\
H_{\alpha\beta\mu\nu} =  \frac{1}{(4\pi s)^{d/2}}& \,\bigg[ \frac{1}{4s^2} \Big( g_{\alpha\mu} g_{\beta\nu} + g_{\alpha\nu} g_{\beta\mu} + g_{\alpha\beta} g_{\mu\nu} \Big) \Big( 1 + \frac{1}{6} s R \Big) \\
& \, - \frac{1}{12s} \Big(g_{\alpha\mu} R_{\beta\nu} + g_{\alpha\nu} R_{\beta\mu} + g_{\beta\mu} R_{\alpha\nu} + g_{\beta\nu} R_{\alpha\mu} + g_{\alpha\beta} R_{\mu\nu} + g_{\mu\nu} R_{\alpha\beta} \Big) \bigg] \unit \, . 
\end{split}
\ee 
\subsection{Laplace-transforms and threshold functions}
Generically, the argument of the operator traces is given by functions $W(z)$ of the Laplacian. The inverse Laplace transform
\be\label{eq:Laplace}
W(z) \equiv \int_0^\infty \mathrm{d}s \,\widetilde{W}(s) \, e^{-sz} \, 
\ee
allows to relate such traces to the heat kernel via
\be
{\rm Tr}\left[ W(\Delta) \right] = \int_0^\infty {\rm d}s \, \widetilde{W}(s) \, {\rm Tr}\left[e^{-s\Delta} \right] \, . 
\ee
Substituting the early-time expansion of the heat kernel \eqref{eq:early-time} yields
\be
{\rm Tr}\left[ W(\Delta) \right] = \frac{1}{(4\pi)^{d/2}} \sum_{n=0}^\infty Q_{d/2-n}[W] \int \mathrm{d}^dx \sqrt{g} \, {\rm tr}(a_{2n}) \, ,
\ee
with the $Q$-functionals being
\be
Q_n[W] \equiv \int_0^\infty \mathrm{d}s \, s^{-n} \, \widetilde{W}(s) \, . 
\ee
For $n>0$ it is straightforward to express the $Q$-functional in terms of $W(z)$
\be\label{eq:QWrel}
Q_n[W] = \frac{1}{\Gamma(n)} \int_0^\infty {\rm d}z \, z^{n-1} \, W(z) \, , \quad n > 0 \, . 
\ee

The prototypical form of $W(z)$ encountered in the present computation is
\be\label{eq:Wzproto}
W(z) = \frac{1}{Z_k \, (P_k)^{p_s}(P_k + k^2 \, w)^{p_g}} \, \p_t (Z_k R_k)
\ee
where $R_k(z) \equiv k^2 R^{(0)}(z/k^2)$ is the scalar regulator function introduced in \eqref{Rregdef} and $p_s$ and $p_g$ denote the number of scalar and graviton propagators appearing in the trace. In this case, the $Q$-functionals are conveniently expressed in terms of (a slight generalization of) the threshold functions introduced in \cite{Reuter:1996cp}
\be\label{threshold1}
\begin{split}
\Phi^{p_s, p_g}_n(w) \equiv & \, \frac{1}{\Gamma(n)} \int_0^\infty \mathrm{d}z \, z^{n-1} \, \frac{R^{(0)}(z) - z R^{(0)\prime}(z)}{(z + R^{(0)}(z))^{p_s} \, (z + R^{(0)}(z) + w)^{p_g}}	\, , \\
\tilde{\Phi}^{p_s, p_g}_n(w) \equiv & \, \frac{1}{\Gamma(n)} \int_0^\infty \mathrm{d}z \, z^{n-1} \, \frac{R^{(0)}(z)}{(z + R^{(0)}(z))^{p_s} \, (z + R^{(0)}(z) + w)^{p_g}}	\, .
\end{split}
\ee
For the Litim-type regulator \eqref{litimreg} the integrals in \eqref{threshold1} are easily evaluated analytically
\be\label{thresholdeval}
\begin{split}
\Phi^{p_s, p_g}_n(w) = & \, \frac{1}{\Gamma(n+1)} \frac{1}{(1+w)^{p_g}} \, , \quad 
\tilde{\Phi}^{p_s, p_g}_n(w) =  \, \frac{1}{\Gamma(n+2)} \frac{1}{(1+w)^{p_g}} \\
\end{split}
\ee
We note that for the special value $p_s + p_g = n+1$ one has
\be
\Phi^{n+1}_n(0) = \frac{1}{\Gamma(n+1)} \, , \qquad  n \ge 0 \, , 
\ee
which is independent of the choice of regulator function and encodes ``universal'' contributions to the flow. To ease our notation, we define the following linear combination of the dimensionless threshold functions
\be\label{eq:defq}
q_{i;n}^{p_s, p_g}(w) \equiv \Phi^{p_s, p_g}_n(w) - \frac{1}{2} \eta_i \tilde{\Phi}^{p_s, p_g}_n(w) \, , 
\ee
where the argument $i \in \{N,s\}$ specifies the anomalous dimension \eqref{anomdim} appearing in the linear combination. 

%

Substituting \eqref{eq:Wzproto} into \eqref{eq:QWrel} then leads to the useful relation
\be\label{eq:Qmaster}
Q_n\left[\frac{\p_t (Z_k R_k)}{Z_k (P_k)^{p_s} (P_k + k^2 w)^{p_g}} \right] = 2 k^{2n-2p_s - 2 p_g+2} \, q^{p_s, p_g}_n(w) \, , 
\ee
and its generalization
\be\label{eq:Qmaster2}
Q_n\left[z^q \, \frac{\p_t (Z_k R_k)}{Z_k (P_k)^{p_s} (P_k + k^2 w)^{p_g} } \right] = 2 \, \frac{ \Gamma(n+q)}{\Gamma(n)} \, k^{2n+2q-2p_s - 2p_g+2} \, q^{p_s, p_g}_{n+q}(w) \, .
\ee
%

%
	%
	%
%
%
%

\section{Traces}
\label{App.D}

The traces which are needed to be computed within our truncation are collected in this appendix. For the gravitational sector, \eqref{gravity-traces}, they evaluate to
\be\label{gravity-traces-eval}
\begin{split}
T_1 = & \, \frac{5}{8\pi^2} k^4 q^1_2 \, \EuScript{I}_0 + \frac{5}{48 \pi^2} k^2 \, q_{1}^1 \, \EuScript{I}_1 \, , \\
T_2 = & \, - \frac{3}{8\pi^2} k^2 \, q_{2}^2 \, \EuScript{I}_1 \\
& \, + \frac{G_k Z_k k^2 }{\pi} \Big(
6 \left(\tilde{C}_k + D_k \right) k^2 q_{3}^2 \EuScript{I}_2 
- \left(\tilde{C}_k + \frac{14}{3} D_k \right) q_{2}^2 \EuScript{I}_3 
- \left(3 \tilde{C}_k + \frac{10}{3} D_k \right) q_{2}^2 \EuScript{I}_4
\\
&\, - 12 C_k Z_k \, q_{2}^2 \, \EuScript{I}_5  
\Big) \, , \\
T_4 = & \, - \frac{3}{\pi} G_k Z_k k^2  \left(3 \tilde{C}_k + 4 D_k \right) q_3^3 \EuScript{I}_3  
+ 48 G^2_k Z_k^2 \, \Big(q^3_2 - k^2 \tilde{C}_k q^3_3 + k^4 \, ( 5 \tilde{C}_k^2  + 4 \tilde{C}_k D_k \\
& \,- 2 D_k^2 ) q^3_4 \Big)  \, \EuScript{I}_5 \, , \\
T_5 = & \, - \frac{15 G_k Z_k}{\pi} \left( q^4_3 + 2 k^2 ( \tilde{C}_k + 2 D_k) q^4_4 + 4 k^4 (\tilde{C}_k + 2 D_k)^2 q^4_5 \right) \EuScript{I}_3 \\
& \, - 240 G_k^2 Z_k^2 \Big( q^4_3 + 2 k^2 ( \tilde{C}_k + 2 D_k) q^4_4 - 22 k^4 (\tilde{C}_k + 2 D_k)^2 \, q^4_5 
-10 k^6  (\tilde{C}_k + 2 D_k)^2\\
& \, \times (7 \tilde{C}_k + 18 D_k) \,  q^4_6 \Big) \EuScript{I}_5 \, , \\ 
T_6 = & \, T_5 \, , \\
T_7 = & \, 96 \, G_k^2 Z_k^2 \, \Big(
 q^5_4 - 30 k^4 (\tilde{C}_k + 2 D_k)^2  \, q^5_6 + 945 \, k^8 \, (\tilde{C}_k + 2 D_k)^4  \, q^5_8  
\Big) \, \EuScript{I}_5 \, , \\
T_8 = & - \frac{3 G_k Z_k k^2 }{\pi} \left(  (\tilde{C}_k + 4 D_k) q^4_4 - 10 k^4 (\tilde{C}_k + 2 D_k)^2  (\tilde{C}_k + 6 D_k) q^4_6 \right) \EuScript{I}_3 \\
& + \frac{60 \, G_k  Z_k \, k^6}{\pi} \, \tilde{C}_k \, (\tilde{C}_k + 2 D_k)^2 \, q^4_6  \, \EuScript{I}_4 
 - \frac{36 \, G_k Z_k^2 \, k^2}{\pi} \,  C_k  \, \left(q_4^4 - 20 k^4 (\tilde{C}_k + 2 D_k)^2 \, q^4_6  \right) \EuScript{I}_5 \, .
\end{split}
\ee
The scalar traces \eqref{scalar-traces} yield
\be\label{scalar-traces-eval}
\begin{split}
T_9 = & \, \frac{1}{16\pi^2} k^4 q^1_2 \, \EuScript{I}_0 +  \frac{1}{96 \pi^2} \, k^2 \, q^1_1 \, \EuScript{I}_1 \, , \\
T_{10} = & \, - \frac{1}{16\pi^2} k^4 (\tilde{C}_k + 4 D_k) \, q^2_3 \, \EuScript{I}_1 - \frac{3}{4\pi^2} \, C_k Z_k \, k^4 \, q^2_3 \, \EuScript{I}_2 - \frac{1}{12 \pi^2} \, C_k Z_k k^2 q^2_2 (\EuScript{I}_3 - \EuScript{I}_4) \,  , \\
T_{12} = & \frac{1}{\pi^2} C_k \, Z_k \, k^4 \,  q^3_4 \left( (2 \tilde{C}_k + 9 D_k) \EuScript{I}_3 +  \tilde{C}_k \EuScript{I}_4 + 15 Z_k C_k \, \EuScript{I}_5 \right) \\
T_{13} = & \, - \frac{3}{\pi} \, G_k k^2 Z_k \Big( (\tilde{C}_k + 4 D_k) q^4_4
 -10  k^4 (\tilde{C}_k + 2 D_k)^2 (\tilde{C}_k +6 D_k) \, q^4_6 \Big) \EuScript{I}_3 \\
& + \frac{60}{\pi}  G_k k^6 Z_k \tilde{C}_k  (\tilde{C}_k + 2 D_k)^2 \, q^4_6 \, \EuScript{I}_4 \\
& - \frac{36}{\pi} {C}_k G_k k^2 \, Z_k^2 \, \Big( q^4_4 - 20 k^4 ( \tilde{C}_k + 2 D_k)^2 \, q^4_6 \Big) \EuScript{I}_5 \, , \\ 
T_{14} = & \, T_{13} \, , \\
T_{15} = &\,   T_7 , \\
T_{16} = & \, - \frac{3 G_k Z_k}{2\pi} \Big( q^4_3 + 2 k^2 (\tilde{C}_k + 2 D_k) \, q^4_4 +4 k^4 (\tilde{C}_k + 2 \tilde{D}_k)^2 \, q^4_5 \Big) \EuScript{I}_3 \\
& \, - 24 G_k^2 Z_k^2 \Big( q^4_3 + 2 k^2 (\tilde{C}_k + 2 D_k) \, q^4_4 - 22 k^4( \tilde{C}_k + 2 D_k)^2 \, q^4_5 
\\ & \qquad \qquad \quad
- 10 k^6 (\tilde{C}_k + 2  D_k)^2 \, ( 7 \tilde{C}_k +18 D_k) \, q^4_6  \Big) \, \EuScript{I}_5 \, . 
\end{split}
\ee
Finally, the Faddeev-Popov ghost sector \eqref{ghost-traces} contributes with
\be\label{ghost-traces-eval}
\begin{split}
	T_{17} = & \,- \frac{1}{2 \pi^2} k^4 \, \Phi^1_2 \, \EuScript{I}_0 - \frac{1}{12 \pi^2} \, k^2 \, \Phi^1_1 \, \EuScript{I}_1 \, , \\
	T_{18} = & - \frac{1}{8\pi^2} \, k^2 \, \Phi^2_2 \, \EuScript{I}_1 \, . 
\end{split}
\ee
Here all threshold functions are evaluated at zero argument.

The most involved part of the computation is the evaluation of the traces $T_3$ and $T_{11}$ containing the insertion of two off-diagonal vertices only. These traces naturally split into three contributions
\be
T_3 = T_3^{\rm commuting} + T_3^{\rm symmetrization} + T_3^{\rm scalar-symmetrization} + T_3^{\rm commutator} \, , 
\ee
and identical for $T_{11}$. By definition $T_3^{\rm commuting}$ captures all contributions from the trace where all covariant derivatives are taken to be commuting. In addition, there are three types of commutator contributions. $T_3^{\rm symmetrization}$ captures the curvature terms originating from symmetrizing the covariant dertivatives acting on the off-diagonal heat-kernel (cf.\ \eqref{eq:Hdef}) while $ T_3^{\rm scalar-symmetrization}$ captures the curvature terms arising in the process of symmetrizing covariant derivatives acting on the scalar field. Moreover, there are additional contributions arising from commuting the covariant derivatives contained in the ``left'' vertex through the propagator, so that all functions containing the Laplacian $\Delta$ can be combined into a single function $W(\Delta)$ which is then acted upon by all uncontracted derivatives. The explicit computation gives
\be
\begin{split}
	 T_{11}^{\rm commuting} = & \,  \frac{2 \, G_k \, Z_k \, k^2}{\pi} \left(
   q^3_3 - 9 k^4  \, (\tilde{C}_k + 2 D_k)^2 \, q^3_5 
	 \right) \, \EuScript{I}_2 \\ & 
	 + \frac{G_k Z_k}{6\pi} \left( q^3_2 + 12 \, k^2 \,  ( \tilde{C}_k + 4 D_k) \, q^3_3 + 3 k^4  \, ( \tilde{C}_k + 2 D_k) \, ( \tilde{C}_k + 10 D_k) \,  q^3_4 \right) \EuScript{I}_3 \\ & 
	 + \frac{G_k Z_k k^2}{2\pi} \left( (21 \tilde{C}_k + 8 D_k) q^3_3 + k^2 \,  ( \tilde{C}_k + 2 D_k)  \,  (23 \tilde{C}_k + 20 D_k)  \,  q^3_4 \right) \EuScript{I}_4 \\ & 
	 + \frac{24 C_k \, G_k Z_k^2 \, k^2}{\pi} \left( q^3_3 + k^2 ( \tilde{C}_k + 2 D_k ) \,  q^3_4 \right) \EuScript{I}_5 \, , 
\end{split}
\ee
together with $T_3^{\rm commuting} = T_{11}^{\rm commuting}$. The contributions arising from the symmetrization of the covariant derivatives acting on the off-diagonal heat-kernel yields
\be
\begin{split}
T_3^{\rm symmetrization} = & \, \frac{G_k \, Z_k}{2\pi} \left(q^3_2 + 2 k^2 (\tilde{C}_k + 2 D_k) q^3_3 + 6 k^4 (\tilde{C}_k + 2 D_k)^2 q^3_4  \right) \, \EuScript{I}_3 \\
 & \, + \frac{G_k \, Z_k}{\pi} \left(q^3_2 - 3 k^2 (\tilde{C}_k - 4 D_k) q^3_3 + 3 k^4 \tilde{C}_k (\tilde{C}_k + 2 D_k) q^3_4 \right) \, \EuScript{I}_4 \, , 
\end{split}
\ee
and
\be
\begin{split}
	T_{11}^{\rm symmetrization} = & \, - \frac{G_k \, Z_k k^4}{2 \pi} (\tilde{C}_k + 2 D_k)^2 \, q^3_4 \, \EuScript{I}_3 \, \\
	& \, - \frac{G_k \, Z_k \, k^2}{\pi} \left( 2 (\tilde{C}_k - D_k) q^3_3 + k^2 \, (7 \tilde{C}_k - 16 D_k) (\tilde{C}_k + 2 D_k) \, q^3_4 \right) \, \EuScript{I}_4 . 
\end{split}
\ee
From the symmetrization of derivatives acting on the scalar field, we obtain
\be
T_3^{\rm scalar-symmetrization} = \frac{G_k Z_k k^2}{2\pi} \left(  (5 \tilde{C}_k - 2 D_k ) q^3_3 - k^2 \, (7 \tilde{C}_k - 22 D_k) (\tilde{C}_k + 2 D_k) \, q^3_4 \right) \EuScript{I}_4 \, , 
\ee
and
\be
T_{11}^{\rm scalar-symmetrization} = - \frac{G_k Z_k k^2}{2\pi} \left(  ( \tilde{C}_k -10 D_k ) q^3_3 - k^2 \, (11 \tilde{C}_k - 14 D_k) (\tilde{C}_k + 2 D_k) \, q^3_4 \right) \EuScript{I}_4 \, .
\ee
Notably, the threshold functions in $T_3^{\rm commutator} $ and $T_{11}^{\rm commutator}$  vanish when evaluated for the Litim regulator.

\newpage
\bibliography{SSTwAS}

\end{document}